\documentclass[amsmath]{JHEP3}

\usepackage{cite}
\usepackage{epsfig}

\def\lsi{\raise0.3ex\hbox{$<$\kern-0.75em\raise-1.1ex\hbox{$\sim$}}}
\def\gsi{\raise0.3ex\hbox{$>$\kern-0.75em\raise-1.1ex\hbox{$\sim$}}}

\newcommand{\gsim}{\mathop{\gsi}}

\title{Equation of state at finite temperature and chemical potential,
lattice QCD results}

\author{F.~Csikor$^a$, G.I.~Egri$^a$, Z.~Fodor$^a$,S.D.~Katz$^b
\thanks{ On leave from Inst. Theor. Phys., E\"otv\"os Univ.}$, 
K.K.~Szab\'o$^a$,
A.I.~T\'oth$^a$ \\ $^a$Institute for Theoretical Physics, E\"otv\"os
University,\\ P\'azm\'any P. 1/A, H-1117 Budapest, Hungary.\\ $^b$Deutsches
Elektronen-Synchrotron DESY, Notkestr. 85, D-22607, Hamburg, Germany}

\date{\today}

\abstract{ We present an $N_t=4$ lattice study for the equation of
state of ~$2+1$ flavour staggered, dynamical QCD at finite temperature
and chemical potential.  We use the overlap improving multi-parameter
reweighting technique to extend the equation of state for
non-vanishing chemical potentials.  The results are obtained on the
line of constant physics and our physical parameters extend in
temperature and baryon  chemical potential upto $\approx 500-600$~MeV. }

\preprint{ITB-Budapest 605\\WUB 04-01}
\begin{document}

\section{Introduction}
 
QCD at finite temperature ($T$) and/or chemical potential ($\mu$) is
of fundamental importance, since it describes relevant features of
particle physics in the early universe, in neutron stars and in heavy
ion collisions (for a clear introduction see \cite{Wilczek:1999ym}).
QCD is asymptotically free, thus its high $T$ and high density phases
are dominated by partons (quarks and gluons) as degrees of freedom
rather than hadrons. In this quark-gluon plasma (QGP) phase the
symmetries of QCD are restored. In addition, recently a particularly
interesting, rich phase structure has been conjectured for QCD at
finite $T$ and $\mu$
\cite{Alford:1998zt,Alford:1999mk,Rapp:1998zu,Rajagopal:2000wf}.

Much effort has been devoted recently to heavy ion collisions at CERN
and Brookhaven in order to experimentally detect the quark-gluon
plasma. Clearly, a theoretical understanding of the underlying physics
is of extreme importance.  Extensive lattice QCD calculations were
carried out to give first principle answers e.g. for the
transition\footnote{We use the expression ``transition'' if we do not
want to specify whether we deal with a phase transition or a
crossover.} temperature ($T_c$) and for the equation of state
(EoS). For recent reviews see
\cite{Ukawa:1997ps,Laermann:1998pf,Karsch:2000vy,Ejiri:2001bw,
Kogut:2002kk,Fodor:2002sd,Laermann:2003cv,Muroya:2003qs,Katz:2003up}.
Unfortunately, all these results are at vanishing chemical potential,
whereas the experiments are carried out at non-vanishing $\mu$ values.

Thus, the main goal of the present paper
is to determine the EoS at finite temperature and chemical potential.

QCD at finite $\mu$ can be formulated on the lattice
\cite{Hasenfratz:1983ba,Kogut:1983ia}; however, standard Monte-Carlo
techniques cannot be used at $\mu \neq 0$. The reason is that for
non-vanishing real $\mu$ the functional measure -- thus, the
determinant of the Euclidean Dirac operator -- is complex. This fact
spoils any Monte-Carlo technique based on importance sampling.
Several proposals were studied to solve the problem.  Unfortunately,
none of them was able to give the EoS at non-vanishing $\mu$.

In a recent paper two of us proposed a new method, the so-called
overlap improving multi-parameter reweighting technique
\cite{Fodor:2001au}, to study lattice QCD and give the phase boundary
at finite $T$ and $\mu$. The idea was to produce an ensemble of QCD
configurations at $\mu$=0 and at $T=T_c$.  Then the Boltzmann weights
of these configurations at $\mu \neq 0$ and at $T$ lowered to the
transition temperatures were determined at this non-vanishing $\mu$
using \cite{Ferrenberg:1988yz,Ferrenberg:1989ui}.  Since transition
configurations were reweighted to transition configurations a much
better overlap was observed than by reweighting pure hadronic
configurations to transition ones
\cite{Barbour:1998ej}. We also emphasized that for small
$\mu$ the technique works for temperatures both below and above the
transition temperature.  We generalized the overlap improving multi-parameter
reweighting method to arbitrary number of staggered quarks and applied
it to the $n_f=2+1$ case \cite{Fodor:2002pe}.  Based on the volume ($V$)
dependence of the Lee-Yang zeros of the partition function we
determined the endpoint
\footnote{The same combination of the multi-parameter reweighting and
the Lee-Yang technique was successfully used on quite large lattices
(upto spatial volumes of $60^3$) to locate the endpoint of the
electroweak phase transition \cite{Csikor:1998eu,Aoki:1999fi}.}  of
QCD with semi-realistic masses on $N_t=4$ lattices. We obtained
$\approx 160$~MeV  for the endpoint temperature and $\approx700$~MeV 
for the endpoint baryonic chemical potential.  
Note that using a Taylor expansion
around $\mu=0$, $T\neq 0$ for small chemical potentials could be also
seen as a variant of the multi-parameter reweighting method, which can
be used to determine hadron masses \cite{Choe:2002mt}, thermal
properties \cite{Allton:2002zi} and even to obtain the EoS as was done in
\cite{Karsch:eos} for two flavours. The Taylor expansion method was also used
to evaluate the pressure on quenched configurations~\cite{Gavai:2003mf}.
Recently, simulations
at imaginary chemical potentials and analytic continuation were also used to
determine the phase boundary on the $\mu-T$ plane for 2,3 and 4 flavour
staggered QCD \cite{deForcrand:2002ci,deForcrand:2003hx,D'Elia:2002gd}.

In this paper we suggest a technique by which the EoS is determined
on the line of constant physics (LCP) \footnote{The importance of using 
LCP's is explained in Section 3 below.}.  An LCP can be defined
\footnote{Our first choice for an LCP is given by the bare masses,
later we determine the LCP using renormalized quantites.}  by a fixed
ratio of the strange quark mass ($m_s$) and light quark masses
($m_{ud}$) to the $\mu$=0 transition temperature ($T_c$).  The $\mu=0$
LCP results are compared with earlier ``non-LCP techniques''
(\cite{Bernard:1997cs,Engels:1997ag,Karsch:2000ps}) ,
which calculate the EoS at fixed $m_q a$ (in this approach the
increase of the temperature -- thus the decrease of the lattice
spacings ``$a$'' -- results in an increase of the quark mass). We comment
on the differences.  Our parameter choice corresponds approximately
to the physical strange quark mass. However, the ratio of the pion
mass ($m_\pi$) and the mass of the rho meson ($m_\rho$) is around $0.5
- 0.75$, which is roughly 3 times larger than its physical value.
 The
temperature dependence of the EoS is studied in a wide range.  In our
lattice analysis we use $2+1$ flavour QCD with dynamical staggered
quarks.  The determination of the equation of state at finite chemical
potential requires several observables (e.g. expectation values of the
plaquette $\langle Pl \rangle$, or the chiral condensate $\langle {\bar
\Psi}\Psi\rangle$) at non-vanishing $\mu$ values. In order to obtain
these quantities we use the overlap improving multi-parameter
reweighting technique of Ref. \cite{Fodor:2001au}. We employ the
integral method \cite{Engels:1990vr} to calculate the pressure.  The
energy density ($\epsilon$) can be obtained by combining the results
on the pressure and on the ``interaction measure''
($\epsilon-3p$). Using the pressure, one can directly determine the
quark number density ($n$).

The paper is organized as follows. In Section 2 we summarize the
lattice parameters.  In Section 3 we present the technique by which
the lines of constant physics can be determined. Section 4 presents the
equation of state at vanishing chemical potential. Sections 5 and 6 deal
with the question how to reweight into the region of $\mu\neq 0$ and 
how to estimate the error of the reweighted quantities. 
In Section 7 we give the equation of state for non-vanishing chemical
potential and temperature. Those who are not interested in the details
of the lattice techniques should simply omit Sections 2--6 and jump to
Section 7, or refer to \cite{Fodor:rovid} or \cite{Fodor:rovid2}.
Finally, Section 8 contains a summary and the conclusions.

\section{Lattice parameters}

In this paper we use $2+1$ flavour dynamical QCD with unimproved
staggered action.  We study the system at several gauge couplings and
masses.  Simulations are done for the equation of state along two
different lines of constant physics and at 14 different temperatures.
Following the experiences of previous studies on the EoS the
simulation points are distributed in a way that they are denser around
the transition temperature than elsewhere.  The temperature range
spans up to $3 T_c$.  In physical units our parameters correspond
to pion to rho mass ratio of $m_\pi/m_\rho \approx 0.5 - 0.75$ and
lattice spacings of $a \approx 0.12-0.35$~fm. See the next section for
more details on the measured quantities (masses, string tension:
$\sqrt{\sigma}$ and potential scale: $R_0,R_1$) along the lines of
constant physics.

The finite temperature contributions to the EoS are obtained on
$4\cdot 8^3$, $4\cdot 10^3$ and $4\cdot 12^3$ lattices, which can be
used to extrapolate into the thermodynamical limit (we usually call
them hot lattices).  On these lattices we determine not only the usual
observables (plaquette, Polyakov line, chiral condensates) but also
the determinant of the fermion matrix and the baryon density ($n_B$)
at finite $\mu$. $10000-20000$ trajectories are simulated at each bare
parameter set. Plaquettes, Polyakov lines and the chiral condensates
are measured at each trajectory whereas the CPU demanding
determinants and related quantities are evaulated at every 30
trajectories. For our parameters the CPU time used for the production
of configurations is of the same order of magnitude than the CPU time
used for calculating the determinants.

Zero temperature runs are done on $12^4$ and $14^3\cdot 24$ lattices
(we usually call them cold lattices).  $1500-2500$ trajectories are
simulated at each bare parameter set.  Hadron propagators and Wilson
loops are measured at every 10 trajectories. Masses and the static
potential are determined by correlated fits. The proper fitting
interval is chosen to obtain minimal $\chi^2$/d.o.f., and/or
$\chi^2$/d.o.f.$\approx$1. In order to avoid instability when
inverting the correlation matrix we use the smeared eigenvalue
technique \cite{Michael:1994sz}.

\begin{figure}
\begin{center}
\epsfig{file=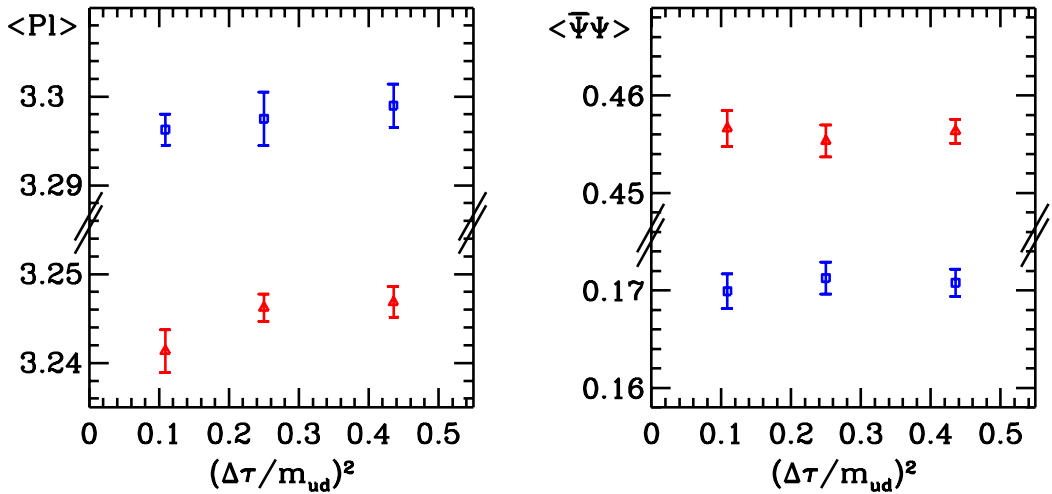,height=7.5cm,width=15cm}
\end{center}
\caption{\label{step-size} (a) The variation of the plaquette as a
function of the step-size squared at $\beta = 5.4$ and $m_qa =0.064$. The
boxes come from the cold and the triangles from the hot lattices. The
cold system has a stronger step-size dependence. As it can be seen
step-sizes smaller than $m_{ud}/2$ result in smaller systematic
uncertainties than the statistical error.  (b) The same for the chiral
condensate of the light quarks. Their step-size dependence is
negligible.}
\end{figure}

Our simulations employ the standard R-algorithm \cite{Gottlieb:mq}.
The length of one trajectory is unity.  The step-size dependence of
different relevant quantities (expectation values of the plaquette
$\langle Pl \rangle$, or chiral condensate $\langle {\bar
\Psi}\Psi\rangle$) was already analysed in the literature (see
e.g. \cite{Blum:1994zf,Bernard:1997cs}).  We also studied the
systematic uncertainties associated with the step-size dependence of
the results. Similarly to other groups we  found that these
effects are much more pronounced for cold lattices. The most sensitive
quantity is the plaquette difference between hot and cold lattices.
This difference  can be of the order of the step-size error.
Figure \ref{step-size} shows the plaquette and the chiral condensate
of the light quarks as a function of the step-size squared, which is
the leading error in the R-algorithm \cite{Gottlieb:mq}.  Based on
these experiences the step-size is chosen to be $m_{ud}/2$. The errors
that are introduced this way are 0.2 \% for zero temperature lattices
and 0.1 \% for finite temperature ones.

Statistical uncertainities are determined by jackknife analysis with
20-50 jackknife samples.

Since we usually move along the line of constant physics by changing
the lattice spacing $a$ and keeping the masses fixed we will
explicitely write out the lattice spacing $a$ in our formulas.  In
this paper we study lattices with isotropic couplings.  In this case
the lattice spacing is the same in the temporal and spatial
directions.  We write $\mu_B$ for the baryonic chemical potential,
whereas for the quark chemical potential ($u,d$ quarks) we use the
notation $\mu$.  Similarly, the baryon density is denoted by $n_B $
and the light quark density by $n$.

\section{Lines of constant physics (LCP) at $\mu=0$}

In this section we discuss the role of LCP when determining the EoS in
pure gauge theory and in dynamical QCD. After that we determine the
lines of constant physics, along which our simulations are done.

The EoS has been determined for pure gauge (quenched) theory on the
lattice already in the early years of lattice QCD. 
 Continuum extrapolations have been made by using several
lattice actions. The results are in agreement within errorbars
\cite{Boyd:1996bx,Okamoto:1999hi,Beinlich:1999ia}.  Results obtained
by the integral method and the derivative method were also compared
\cite{Ejiri:1998xd,Engels:2000tk}.

In order to detetermine the temperature $T=1/[N_ta(\beta)]$ of the
pure gauge theory, we have to compute the lattice spacing ($a$) as a
function of the gauge coupling ($\beta$). Note that when changing the
coupling of the pure gauge system one automatically remains (upto
scaling violations) on the line of constant physics.  The situation is
quite different in full QCD. In the $d$ dimensional space of the bare
parameters one defines $d$ appropriately chosen quantities. The LCP is
given by $d-1$ constraints and it is parametrized by a non-constrained
combination of the above quantities.  For the $2+1$ flavour staggered
action we have three bare parameters ($\beta$, and two masses,
$m_{ud},m_s$). Thus, we need two constraints. There are several
possibilities for these constraints and consequently there are many
ways to define an LCP.  A convenient choice for two of the three
quantities can be the bare quark masses ($m_{ud}$ and $m_s$). A more
physical possibility is to use the pion and kaon masses
($m_\pi,m_K$). There are several options for the third quantity.  It
can be the rho mass ($m_\rho$), the string tension ($\sqrt{\sigma}$),
the $R_0,R_1$ scales of the potential (\cite{Sommer:1994ce}, and also
\cite{Bernard:2001av}) or the transition temperature. Fixing the
ratios of the third quantitiy to the first two gives two constraints
and the third quantity in lattice units fixes the scale along the LCP.

In our analysis first we use the bare quark masses ($m_{ud}$ and
$m_s$) and the transition temperature to define an LCP. In this paper
we use two\footnote{As we will see later, two LCP's are needed for the
determination of the EoS at finite chemical potential.} 
LCP's (LCP$_1$ and LCP$_2$).  The
conditions
\begin{eqnarray}\label{constraint}
\begin{array}{c}
m_{ud}=0.48T_c =0.48/(N_ta)  \ \ \ \ {\rm and}\ \ \ \ \  
m_s=2.08\cdot m_{ud}\\
m_{ud}=0.384T_c=0.384/(N_ta) \ \ \ \ {\rm and}\ \ \ \ \  
m_s=2.08\cdot m_{ud}
\end{array}
\end{eqnarray}
are used as the constraints for LCP$_1$ and LCP$_2$,
respectively\footnote{Our light quark masses are
heavier than the physical ones, whereas the strange quark mass is at
its approximate physical value.}.  For both LCP's we determined four
different transition couplings ($\beta_c$) by susceptibility peaks on
$N_t=1/(T_ca)=4$, 6, 8 and 10 lattices with spatial extensions
$N_s\gsim 3N_t/2$ and quark masses given by eq.~(\ref{constraint}).
The quark masses or the transition gauge couplings can be used to
parametrize the LCP's.

Note that $m_s/m_{ud}$ is the same for both of our LCP's. This
relationship does not change when we interpolate between the two LCP's
or perform reweighting in some parameters. Thus, for transparency we
usually do not specify the quark flavours when discussing the physical
scale. We simply speak about ``quark mass parameter'' or
``quark mass''.  The flavour dependence is indicated explicitely only
when necessary.

It is instructive to define LCP's by using renormalized quantities
(LCP$^*$'s, namely LCP$_1^*$ and LCP$_2^*$) and to test scaling
violation. One can calculate $m_\rho,m_\pi$, $R_0$ and $\sigma$ on $T=0$
lattices ($V = 14^3\cdot 24$) using the bare parameters obtained for
LCP$_1$ and LCP$_2$. By measuring these quantities at somewhat
different quark masses gives the possibility to define by linear
interpolations the LCP$^*$'s. Our results for LCP$_2$ and LCP$_2^*$
are summarized in Table \ref{scaling} and Table \ref{scaling2}. The
definition of LCP$_2^*$ was $m_\pi/m_\rho \approx 0.626$. As it can be seen
different definitions of the scales deviate from each other only by
$\approx$ 8 \%. Table \ref{scaling} and Table \ref{scaling2} contains
the $T=0$ results for the ``non-LCP approach'', too (see next section).


\begin{table}[htb]
\begin{center}
\begin{tabular}{|c|c|c||c|c|c|c|c|}
\hline
$N_t$  & $\beta$  & $m a$ & $\sqrt{\sigma} a$ & $R_0/a$  & $R_1/a$   & $m_\rho
a$ & $m_\pi a$ \\
\hline \hline
4 & 5.271 & 0.096 & 0.5889(9)& 2.02(1)& 1.483(2)& 1.421(1)& 0.78514(6) \\
6 &5.4& 0.064 & 0.3686(6) & 3.19(5) & 2.34(2)& 1.048(1)& 0.6805(1) \\
8 &5.5 & 0.048& 0.2697(5) & 3.96(9) & 3.64(10)&  0.778(4) & 0.5595(13) \\
10 &5.58 & 0.0384&  0.2358(4) & 4.44(5)& 3.71(5)& 0.637(4)& 0.480(3) \\
\hline \hline
4 &5.271 & 0.1412 & 0.350(1)& 2.94(4) & 2.71(3) & 1.511(7)&  0.9282(3)
\\
6 &5.4 & 0.067& 0.426(2)& 2.92(7)& 2.21(9)& 1.076(3)& 0.693(10) \\
8 &5.5 & 0.043& 0.344(2)& 3.27(28) & 2.50(8) & 0.933(15)& 0.566(1) \\
10 &5.58 &0.0304& 0.347(2)& 2.95(4)& 2.72(3)& 0.86(3)& 0.469(2) \\
\hline \hline
3.2 &5.19 & 0.096& 0.6846(11)& 1.748(3)& 1.327(2)& 1.450(5)& 0.7677(6) \\
4 & 5.27 &0.096& 0.5899(9)& 2.02(1)& 1.483(2)& 1.421(1) & 0.78514(6) \\
4.8 &5.36 & 0.096& 0.4613(6) &  2.55(13)& 1.81(1)& 1.314(3) &
0.8045(3) \\
6.21 &5.5 & 0.096& 0.3239(30)& 3.85(7) & 2.63(1)& 1.081(1)& 0.803(1) \\
7 & 5.58 & 0.096& 0.2716(10)& 4.8(1)& 2.94(9)& 0.9513(8)& 0.777(2) \\
\hline
\end{tabular}
\caption{\label{scaling} Different quantities along LCP$_2$ (upper
part) and the ``non-LCP approach'' (lower part).  The middle part
shows new simulation points that were used to determine LCP$_2^*$.  As
usual, the numbers in the parenthesis indicate the estimated
statistical errors of the last digits. The effective $N_t$ values for
the ``non-LCP approach'' were calculated as follows. We determined
four different critical couplings on $N_t$=4,6,8 and 10 lattices,
while the quark mass was fixed at $ma$=0.096. Effective $N_t$-s were
determined by interpolating between and slightly extrapolating from
these points.}
\end{center}
\end{table}

\begin{table}[htb]
\begin{center}
\begin{tabular}{|c|c|c||c|c|c|c|c|}
\hline
$N_t$  & $\beta$  & $m a$ & $\sqrt{\sigma}R_0$ & $m_\rho R_0$
&$\sqrt{\sigma}R_1$  & $m_\pi/m_\rho$\\
\hline \hline
4 & 5.271 & 0.096& 1.190(8)& 2.87(2) & 0.874(3)&0.552(4) \\
6 &5.4& 0.064& 1.176(2)& 3.34(6) & 0.860(9) &0.649(1) \\
8 &5.5 & 0.048 & 1.06(3)& 3.08(9)& 0.98(3)&0.719(5) \\
10 &5.58 & 0.0384 & 1.05(1)& 2.83(5) & 0.87(1) &0.75(1) \\
\hline \hline
4 &5.271 & 0.1412 & 1.03(2)& 4.44(8) &  0.95(1)&0.614(4) \\
6 &5.4 & 0.067 & 1.24(4) & 3.14(8)& 0.94(4) &0.644(11) \\
8 &5.5 & 0.043 & 1.12(10)& 3.05(31)& 0.86(3)&0.60(1) \\
10 &5.58 & 0.0304 & 1.02(2)& 2.52(12)& 0.94(2)&0.55(2)\\
\hline \hline
3.2 &5.19 & 0.096 & 1.197(3)& 2.53(1)& 0.908(3)&0.529(3) \\
4 & 5.271 & 0.096& 1.190(8)& 2.87(2)& 0.874(3)&0.552(4) \\
4.8 &5.36 & 0.096& 1.18(6) & 3.35(18)& 0.835(6)&0.612(3) \\
6.21 &5.5 & 0.096& 1.25(3)& 4.16(8)& 0.852(11)&0.743(2) \\
7 & 5.58 & 0.096& 1.30(3)& 4.57(10)& 0.80(3)&0.817(3) \\
\hline
\end{tabular}
\caption{\label{scaling2} Dimensionless combinations along LCP$_2$
(upper part) and the ``non-LCP approach'' (lower part).  The middle
part shows new simulation points that were used to determine
LCP$_2^*$. Quark mass dependent dimensionless quantities change more
in the ``non-LCP approach''.  }
\end{center}
\end{table}

\begin{figure}
\begin{center}
\epsfig{file=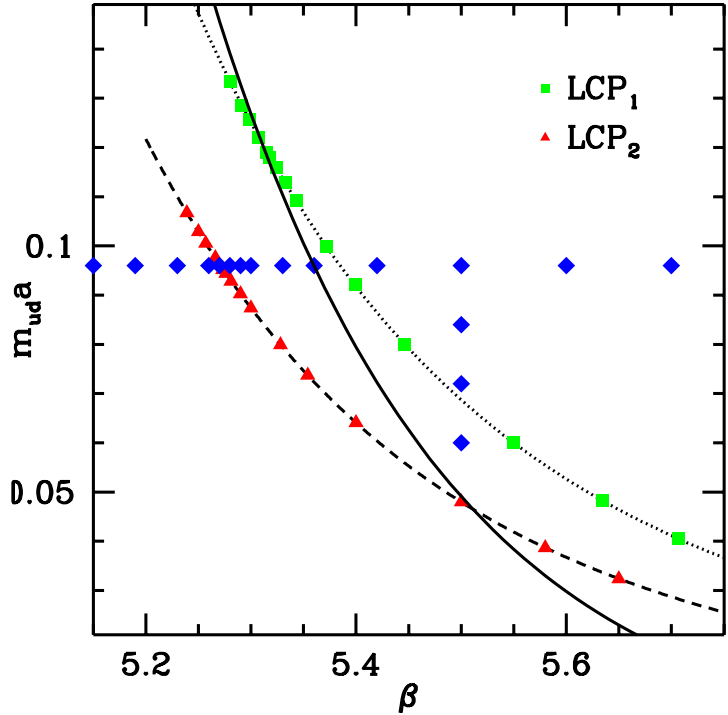,width=10cm}
\end{center}
\caption{\label{LCP} The lines of constant physics (LCP$_1$ and
LCP$_2$) on the $\beta$ vs. $m_{ud}a$ plane.  The strange quark mass
is given by $m_s=2.08 m_{ud}$ for both LCP's. The simulation points are
shown by squares/triangles and connected by dashed/dotted lines for
LCP$_1$/LCP$_2$, respectively. An approximate LCP
obtained by renormalized quantities (LCP$_2^*$, the definition is
$m_\pi/m_\rho \approx 0.626$) is shown by a solid line.  The diamonds
along a horizontal line represent the simulation points in the
``non-LCP'' approach. Additional 3 diamonds in the vertical direction
show the simulation points used to test the path independence of the
integral method (see next Section).  }
\end{figure}

By the finite temperature technique, described above, only a few
points of the LCP's can be obtained. To interpolate $\beta(a)$ between
these points (and extrapolate slightly away from them) we use the
renormalization group inspired ansatz proposed by Allton
\cite{Allton:1996kr}.  Note that not only the quark mass parameter
($m_q a$) can be used as a parameter of the LCP's, but a particularly
illustrative parametrization is obtained by inverting
eq.~(\ref{constraint}) and using $N_t$ as a continuous parameter. (We
use later this $N_t$ parametrization for constructing the LCP on the
$\mu - \beta$ plane by a linear combination of LCP$_1$ and LCP$_2$,
see Section 5.)  Figure \ref{LCP} shows LCP$_1$ and LCP$_2$ with our
simulation points.  A line of constant physics obtained by
renormalized quantities (LCP$_2^*$) and the simulation points in the
``non-LCP'' approach are also shown.  Note that even though the
determination of the LCP$_1$ and LCP$_2$ are done on finite
temperature lattices, the obtained bare parameters are used in the
rest of the paper for $T=0$ and $T \neq 0$ simulations.

\section{Equation of state along the lines of constant physics  (LCP) 
at $\mu=0$}\label{sec_mu0}

In previous studies of the EoS with staggered quark actions, the
pressure and the energy density were determined as functions of the
temperature for fixed value of the bare quark mass $m_qa$   in the 
lattice action \cite{Bernard:1997cs,Engels:1997ag,Karsch:2000ps}. In
these studies a fixed $N_t$ was used (e.g. $N_t = 4$ or 6) at different
temperatures.  Since $T=1/(N_t a)$ the temperature is set by the
lattice spacing which changes with $\beta$.  This convenient, fixed
bare $m_qa$ choice leads to a system which has larger and larger
physical quark masses at decreasing lattice spacings (thus, at
increasing temperatures). Increasing physical quark masses with
increasing temperatures could result in systematic errors of the EoS.

Clearly, instead of this sort of analysis (in the rest of the paper we
refer to it as ``non-LCP approach'') one intends to study the
temperature dependence of a system with fixed physical observables,
therefore on an LCP.

Recently, the EoS was also studied by using an action
with dynamical Wilson quarks \cite{AliKhan:2001ek}. In this case there
is no convenient choice similar to the approach with fixed $m_q a$
(``non-LCP approach'').  The authors determined the lines of constant
physics and the $\beta$ functions. They obtained the EoS along
different LCP's in the range of $m_\pi/m_\rho =0.65-0.95$.

In our analysis we use full QCD with staggered quarks along the LCP
and compare these results with those of the ``non-LCP approach''.

In order to be self-contained in this Section we review the $\mu = 0$
technique to determine the interaction measure ($\epsilon - 3 p$) by
using the non-perturbative $\beta$-function. Then we shortly discuss
the integral method \cite{Engels:1990vr} to determine the pressure.
In practice, the energy density, entropy density or the speed of sound
can be directly deduced from the interaction measure and the pressure
(c.f. $sT = \epsilon + p$ and $c_s^2 = dp/ d\epsilon$).  We review
the basic formulas and emphasize the issues related to the EoS
determination along an LCP.

The energy density and pressure are defined in terms of the
free-energy density ($f$):
\begin{eqnarray}
\epsilon (T)=f-T \frac{\partial{f}}{\partial {T}},\ \ \ \ \ \ \ \ && p(T) = -f.
\end{eqnarray} 
Expressing the free energy in terms of the partition function 
($f=-T/V \log Z =$ \\ $ -T \partial (\log Z)/\partial V$) we have:
\begin{eqnarray} \label{eps_p}
\epsilon (T)=\frac{T^2}{V}\frac{\partial{\log Z}}{\partial {T}}, \ \ \
\ \ \ \ \ && 
p(T) = T\frac{\partial (\log Z)}{\partial V}.
\end{eqnarray} 

As mentioned before, we use the same lattice spacings in the temporal
and spatial directions.  The temperature and volume are connected to
this lattice spacing by
\begin{eqnarray} \label{TV}
T=\frac{1}{a N_t}, && V=a^3 N_s^3.
\end{eqnarray}

By varying the bare parameters one can change the lattice spacing;
however, it is not possible to change $T$ and $V$ independently. This
is the reason why eqs. (\ref{eps_p},\ref{TV}) cannot be used
directly. Instead one determines the interaction measure by the help
of the $\beta$ function and the pressure by using the integral method.

Inspecting eqs.(\ref{eps_p}, \ref{TV}) we see that the interaction
measure $(\epsilon-3p)/T^4$ is directly proportional the total
derivative of $\log Z$ with respect to the lattice spacing:
\begin{equation}\label{interaction}
\frac{\epsilon-3p}{T^4}=-\frac{N_t^3}{N_s^3}a \frac{d (\log Z)}{d a}.
\end{equation}
Here, the derivative with respect to $a$ is defined along the LCP,
which means that only the lattice spacing changes and the physics (in
our case $m_q/T_c$) remains the same. The variation of the lattice
spacing can be controlled by the bare parameters of the action
($\beta$ and $m_q a$), so we can write:
\begin{equation}
\frac{d}{da}=\frac{\partial \beta}{\partial a}\frac{\partial}{\partial \beta} +
\sum_q \frac{\partial (m_q a)}{\partial a}\frac{\partial}{\partial (m_q a)}.
\end{equation}
The partial derivatives with respect to $a$ should be taken along
the LCP. Since the LCP is defined by $m_q/T_c={\rm const.}$, the
partial derivative $\partial (m_qa)/\partial a$ becomes simply $m_q$.
The derivatives of $\log Z$ with respect to $\beta$ and $m_q$ are the
plaquette and $\bar{\Psi}\Psi_q$ averages multiplied by the lattice
volume. We get:
\begin{equation}\label{interaction-measure}
\frac{\epsilon-3p}{T^4}=-N_t^4 a 
\left(\overline{{\rm Pl}} \left. \frac{\partial \beta}{\partial a}\right|_{\rm  LCP}+
\sum_q \overline{\bar{\Psi}\Psi}_{q} m_{q} \right).
\end{equation}

The pressure is the other basic quantity, which is usually determined
by the integral method \cite{Engels:1990vr}.  The pressure is simply
proportional to $\log Z$, however it cannot be measured directly. One
can determine its partial derivatives with respect to the bare
parameters. Thus, we can write:
\begin{equation}\label{integral}
\frac{p}{T^4}=\left[-\frac{N_t^3}{N_s^3}\int^{(\beta,m_q a)}_{(\beta_0,m_{q0} a)}
d (\beta,m_q a) 
\left(\begin{array}{c}
{\partial \log Z}/{\partial \beta} \\
{\partial \log Z}/{\partial (m_q a)}
\end{array} \right )\right]- \frac{p_0}{T^4}.
\end{equation}
Since the integrand is the gradient of $\log Z$, the result is by
definition independent of the integration path. We need the pressure
along the LCP, thus it is convenient to measure the derivatives of
$\log Z$ along the LCP and perform the integration over this line. For
the subtracted vacuum term we used the zero temperature pressure,
i.e. the same integral on $N_{t0}= N_s$ lattices.  The lower limits of
the integrations (indicated by $\beta_0$ and $m_{q0}$) were set
sufficiently below the transition point. By this choice the pressure
becomes independent of the starting point (in other words it vanishes
at vanishing temperature).  In the case of $2+1$ staggered QCD
eq. (\ref{integral}) can be rewritten appropriately and the pressure
is given by
\begin{equation}\label{pmu0}
\frac{p}{T^4}=
-N_t^4\int^{(\beta,m_q a)}_{(\beta_0,m_{q0} a)}
d (\beta,m_{ud} a,m_s a) 
\left(\begin{array}{c}
\langle{\rm Pl}\rangle \\
\langle\bar{\Psi}\Psi_{ud}\rangle \\
\langle\bar{\Psi}\Psi_{s}\rangle
\end{array} \right),
\end{equation}
where we use the following notation for subtracting the vacuum term:
\begin{equation}
\langle  {\cal O}(\beta,m) \rangle= 
{\overline {{\cal {O}}}(\beta,m)}_{T\neq 0}-
{\overline {{\cal O}}(\beta,m)}_{T=0}.
\end{equation}

The integral method was originally introduced for the pure gauge case
for which the integral is one dimensional, it is performed along the
$\beta$ axis. Previous studies for staggered dynamical QCD
\cite{Bernard:1997cs,Engels:1997ag,Karsch:2000ps} used a
one-dimensional parameter space. Note that for full QCD the
integration should be performed along a path in a multi-dimensional
parameter space. In our $2+1$ flavour staggered QCD case the parameter
space is three-dimensional.

According to eq.(\ref{integral}) the gradient of $\log Z$ is 
measured, thus the integral should be independent of the path.  We
explicitely checked this independence by performing the integration
along different paths.  For this purpose we determined the normalised
pressure ($p/T^4$) at $\beta=5.5$, $m_{ud} =0.048$ (this point
corresponds to $T=2 T_c$ with quark masses defined by
eq.(\ref{constraint})).  In the calculation two different integration
paths were used.  In the first case we integrated along the LCP. We
obtained $p/T^4=6.15(5)$. In the second case the integration path
contained two pieces (see the path defined by the diamonds of Figure
\ref{LCP}).  The first piece is a $\beta$ integral at constant $m_q$,
the second one is a $m_{ud}, m_s$ integral (with fixed mass ratio) at
constant $\beta$ value. For the second path we obtained
$p/T^4 = 6.15(4)$. As it can be seen the two different paths give the
same result for the pressure (within the estimated uncertainties).

\begin{figure}
\begin{center}
\epsfig{file=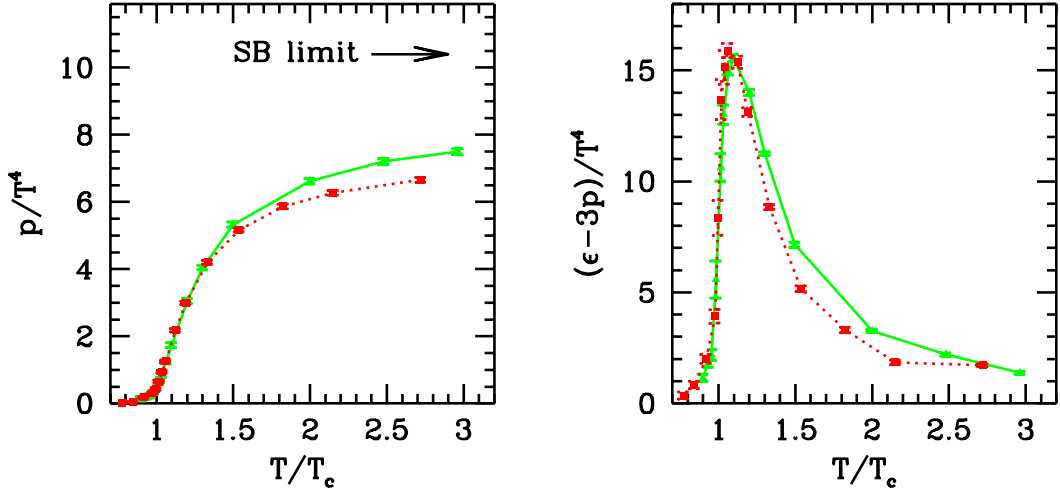,height=7.5cm,width=15.0cm}
\end{center}
\caption{\label{eosmu0} The equation of state at $\mu$=0. (a) The left
panel shows the pressure $p$, as a function of the temperature.  All
quantities are normalised by $T^4$.  The
solid line connects the data points obtained along the LCP$_2$,
whereas the dashed line connects the data points obtained in the
``non-LCP approach'' (see text). The Stefan-Boltzmann limit is also
shown by an arrow for $N_t$=4 lattices.  The EoS along the LCP and the
``non-LCP approach'' differ from each other at high $T$.  (b) The same
for the ``interaction measure'' ($\epsilon$-3p).}
\end{figure}

The EoS is determined along the LCP's and also by using the ``non-LCP
approach''. The results along LCP$_2$ and LCP$_1$ are very
close to each other (the $T=0$ pion masses differ only by 10\% for the
two LCP's; for the $m_\pi/m_\rho$ dependence of the EoS with Wilson
quarks see \cite{AliKhan:2001ek}).  Figure \ref{eosmu0} shows the
EoS at vanishing chemical potential on $N_t =4$ lattices
for LCP$_2$ and for the non-LCP approach.  The pressure and
$\epsilon-3p$ are presented as a function of the temperature. The
parameters of LCP$_2$ and those of the non-LCP approaches coincide at
$T = T_c$.  The Stefan-Boltzmann limit valid for $N_t = 4$ lattices is
also shown.

 It is interesting to compare the pressure difference between LCP and
 non-LCP approaches at finite temperature in case of the ideal and
 interacting gas. In the ideal gas case there is a relative 2.7\% difference
 between the pressures, while this difference is three times larger in the
 interacting case (9.4\%) at $T=2.5 \cdot T_c$. This means that considering
 only the mass-effect of the ideal gas, we would make a significant 
 underestimation for the interacting case.

\section{Reliability of reweighting}

In order to 
determine the EoS for the $\mu \neq 0$ case (when direct simulations 
are not possible) we use the multi-parameter
reweighting proposed in \cite{Fodor:2001au} 
The aim of this section is to point out that the multi-parameter 
reweighting \cite{Fodor:2001au} is reliable. (For another study of 
reweighting see~\cite{Ejiri:2004yw}.)
We also determine its region of 
validity by a suitably estimated error.
 Furthermore we show that in certain cases the multi-parameter 
reweighting is much more reliable than single-parameter techniques 
(Glasgow-type reweighting \cite{Ferrenberg:1988yz}) in the sense 
that one can reach farther regions. On top of this we 
demonstrate that the major requirement to be met by the sample used for 
the reweighting is that it should be as variable as possible. Variability 
here means that it should contain  configurations from the 
various phases or  configurations 
from the different Z(3) sectors in case of imaginary chemical potentials. 
To start with let us briefly review the  multi-parameter reweighting. 

As proposed in \cite{Fodor:2001au} one can identically rewrite the partition 
function in the form:
\begin{eqnarray}\label{multi-parameter}
Z(m,\mu,\beta) =
\int {\cal D}U\exp[-S_{bos}(\beta_0,U)]\det M(m_0,\mu=0,U)\\
\left\{\exp[-S_{bos}(\beta,U)+S_{bos}(\beta_0,U)]
\frac{\det M(m,\mu,U)}{\det M(m_0,\mu=0,U)}\right\},\nonumber
\end{eqnarray}
where $U$ denotes the gauge field links and $M$ is the fermion matrix 
\footnote{For $n_f\neq 4$ staggered dynamical QCD one simply takes 
fractional powers of the fermion determinant.}. The chemical potential 
$\mu$ is included as $\exp(a\mu)$ and $\exp(-a\mu)$ multiplicative 
factors of the forward and backward timelike links, respectively. In 
this approach we treat the terms in the curly bracket as an observable 
-- which is measured on each independent configuration, and can be 
interpreted as a weight -- and the rest as the measure.  Thus the  simulation 
can be performed at $\mu=0$ and at some $\beta_0$ and $m_0$ values 
(Monte-Carlo parameter set). By using the  reweighting 
formula~(\ref{multi-parameter}) 
one obtains the partition function at another set of parameters, 
thus at $\mu\neq 0$, $\beta\neq\beta_0$ or even at $m\neq m_0$ 
(target parameter set).

Expectation values of observables can be determined by the above technique. 
In terms of the weights (i.e. the expression in the curly bracket of 
eq.~(\ref{multi-parameter})) the averages can be determined as:
\begin{equation} \label{multi-parameter2}
{\overline {\cal O}}(\beta,\mu,m)=\frac{\sum \{w(\beta,\mu,m,U)\}
{\cal O}(\beta,\mu,m,U)}{\sum
\{w(\beta,\mu,m,U)\}}.
\end{equation}

It is clear that simulating at a given Monte Carlo parameter set the errors 
increase as we go farther and farther with the target parameter set. 
We  need an error estimate of the reweighted quantities which 
shows how far we 
can reweight in the target parameter space. This way we could determine the 
borderline of the region outside of which the reweighting already fails 
to give reliable predictions. To achieve this we checked  several 
error definitions including the jackknife and the statistical errors   
 described in the papers \cite{Ferrenberg:1995,Newman}. Both 
proved little to draw the limit of the reweighting procedure. The reason 
for this is that neither of them contains the systematic errors occuring 
because of the finite sample size. We note that in \cite{Newman} there 
is an error estimate recommended for taking the finite sample size 
into consideration. 
However,  this estimate  uses an approximation of the 
distribution of 
the sample. An approximation of this kind can not be justified in case of 
staggered dynamical QCD. 

Eventually, with a new technique different from the ones 
mentioned above we succeeded 
to define a reliable error estimate for the reweighted quantities and to draw 
the limit of the reweighting procedure. After defining the new technique we
 show its application on an example. Then we 
demonstrate the problems that crop up while using the jackknife error 
and furthermore show how our new method overcomes these difficulties. 

The steps of the new procedure are as follows. First we assign to each 
and every starting configuration the weight $w$ valid at the chosen target 
parameter set i.e.   
\begin{equation}
w=\exp\left\{\Delta\beta \cdot V\cdot(Pl)+\frac{n_f}{4}[\ln\det M(\mu)-\ln\det M(\mu=0)]\right\}
\label{eq:newweight}
\end{equation} 
where $V$ is the lattice volume. Note that only the $\beta$ and the 
$\mu$ parameters differ from their values at the simulation point. 
We generate a new set of configurations using these weights for 
Metropolis-like accept-reject steps. Let the original and new configuartions
be $\Phi_i$ and $\Phi'_i$ and the corresponding weights (according to 
eq.~(\ref{eq:newweight})) $w_i$ and $w'_i$. The first new configuration 
is $\Phi'_1=\Phi_1$. Then for $i>1$, $\Phi'_i =\Phi_i$ with the 
probability ${\rm max}\{1,\frac{w_i}{w'_{i-1}}\}$ and 
$\Phi'_i=\Phi'_{i-1}$ otherwise. For complex weights we have to take either
the absolute values or the absolute values of the real parts of the weights. 
The $\Phi'_i$ configurations of the new sample 
are  taken with a unit weight to calculate expectation values, 
variances and integrated autocorrelation times \cite{Ferrenberg:1995} 
(the latter grows at every rejection due to the repetition of certain 
configurations). If our initial sample consisted of  a  large number 
of independent configurations then for real weights this method regenerates 
a sample  which is theoretically perfect for the target parameters.
For complex weights the new set of configurations can not be used to 
measure observables, but it can be used for getting an error estimate.
Therefore, in general, we will use 
eq.~(\ref{multi-parameter2}) to get the expectation values and the newly 
generated sample to determine the errors (see later).

We still have to clarify two important points before our new procedure 
is finalised.  The first is the question of thermalization. 
The second question is (in an extreme case):  what happens if there is no 
configuration in the starting sample which would be relevant at the 
target parameters. We solved the first problem  by defining a thermalization 
segment at the beginning of every newly generated sample which we cut 
off from the sample before calculating the expectation values and the 
errors. An obvious assumption is to claim  that the already thermalized 
sample containing valuable 
information starts with the first different configuration right after 
the one with the largest weight. This ensures that  
if there is only one configuration in the initial 
sample which ``counts'' at the target parameters then this information 
will not be lost. The second problem can not be 
 solved perfectly. The problem occurs e.g. when a  
phase transition is very strong, that is the physical quantities change 
significantly during the transition (first order transition) and we would 
like to reweight starting from deep inside in one phase into deep inside 
into the other phase. Then the 
solution can only be a huge sample size (or small lattice size) which 
allows the presence of a few  configurations of the target phase 
in the initial sample. It is, however, difficult to define what a huge 
sample really means in a general, practical   way.

\begin{figure}[ht]
\begin{center}
\includegraphics*[width=7.0cm,bb=20 170 570 700]{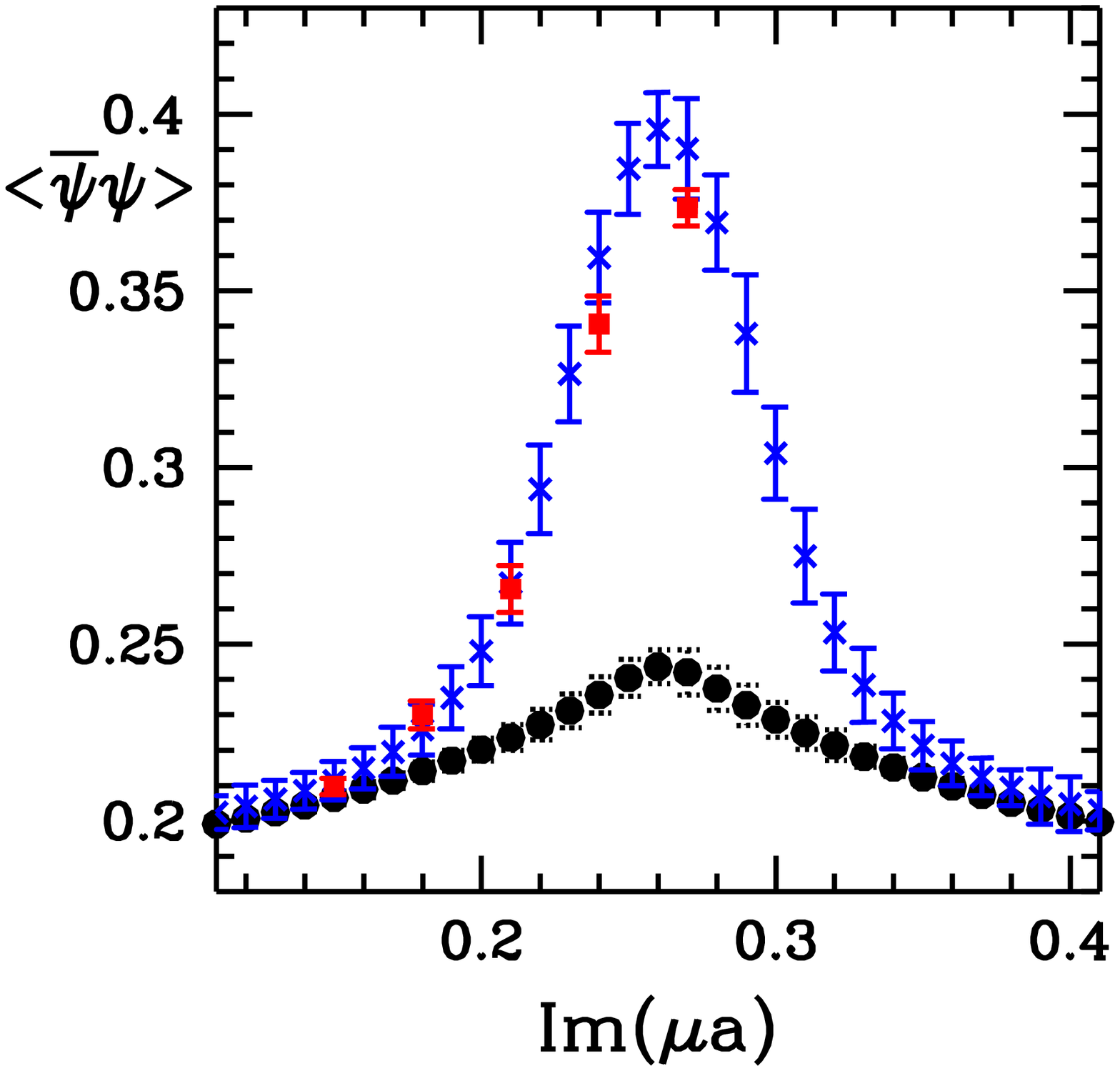}
\includegraphics*[width=7.0cm,bb=20 170 570 700]{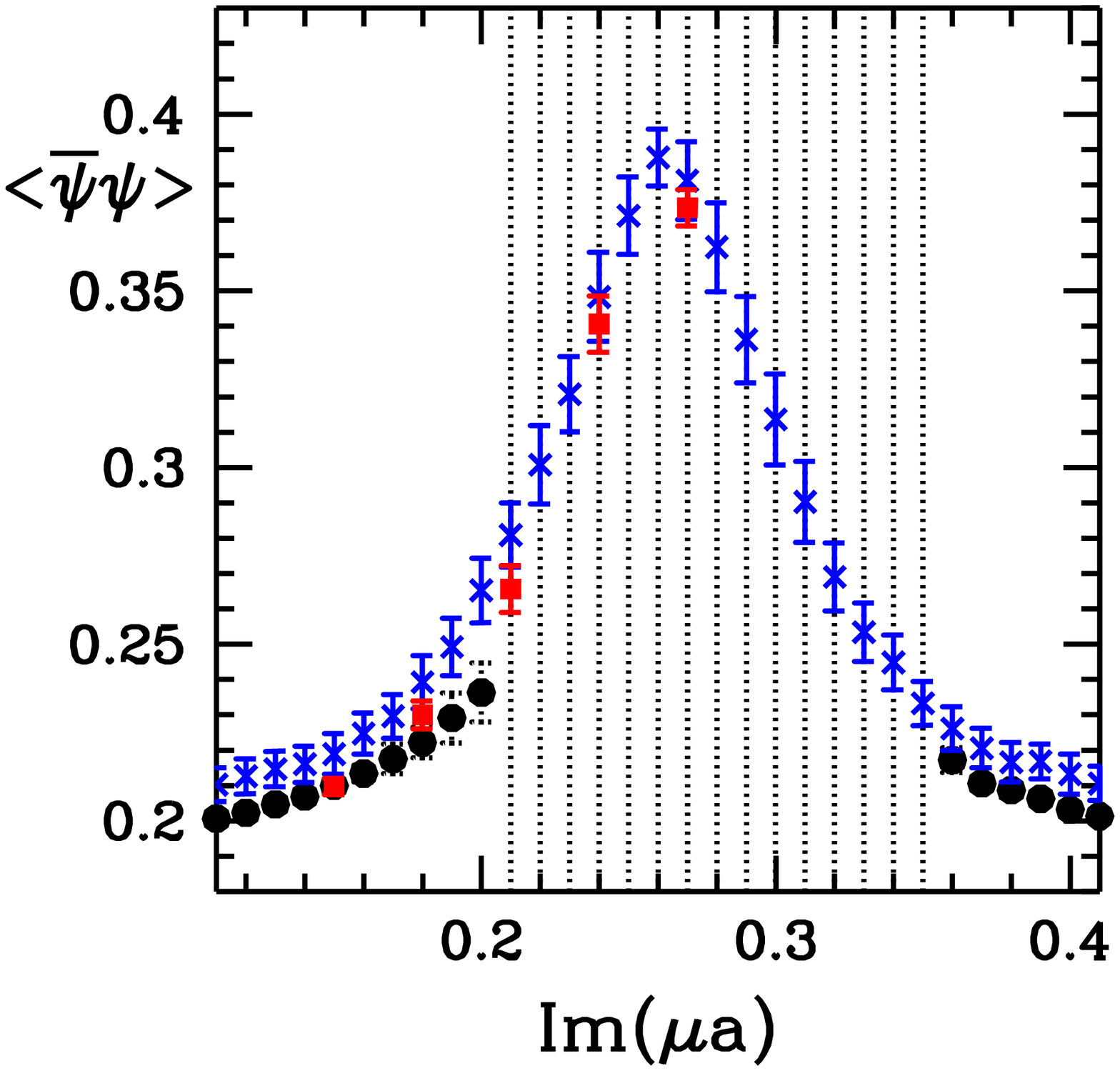}
\end{center}
\caption{\label{fig:metro1} {
(a) The left panel shows $\bar{\psi}\psi$ expectation values 
and error estimates 
of  three different methods. The squares correspond  to direct 
simulations (out of $60000$ configurations). 
The crosses denote the results of the multi-parameter reweighting method 
(out of $\approx 1200$ independent configurations) 
while the circles are the points of the Glasgow-type reweighting (out 
of $\approx 1200$ independent configurations). In case of the latter two methods the error estimates are determined using the new Metropolis-type method.
(b) In the right panel the meaning of the 
symbols are unchanged but  the sample sizes for the reweighting 
techniques are increased to 2500 independent configurations in case of the 
multi-parameter reweighting and to 7000 in case of the Glasgow-type 
reweighting. 
In case of the multi-parameter reweighting the increased sample size only 
decreases the error estimate, maintaining consistency with the direct 
simulation result.  On the other hand 
in case of the Glasgow-type reweighting the fact that the small sample 
(used in the left panel) does
not contain any configuration from the target phase causes systematic
errors in the expectation values and mainly in their uncertainties.
This is because in case of strong phase transitions -- like in our case
-- it is far too difficult to define a reliable thermalization stage
by  virtue of the weights. Increasing the sample size helps in this
situation leading to the results of the right panel. The very large error 
estimates in the Glasgow case are realistic, the prediction does agree within 
errors with the direct simulation result.
}}
\end{figure}

To illustrate the new technique let us take a look at the $n_f=4$ 
flavour case at $m_qa=0.05$ bare quark mass 
on $4\cdot 6^3$ size lattice at imaginary chemical potential. 
Note that for purely imaginary chemical potentials direct simulation 
is possible, therefore it is possible to check the validity of any 
 error estimation method. Direct simulation is thus used as a standard.
Accordingly, we determine the plaquette ($Pl$) and the $\bar{\psi}\psi$ 
expectation values and their uncertainties at $\beta=5.085$ and 
Im$(\mu)\neq 0$, that is at the target, imaginary chemical potential values 
using three different techniques. First technique is direct simulation 
(60000 configurations). The other two methods are based on reweighting 
Im$(\mu)=0$ data.  We carried out simulations
at Im$(\mu)=0$ in the phase transition point, i.e. at 
$\beta=5.04$ ($\approx 2500$ independent configurations) and at $\beta=5.085$ 
($\approx 7000$ independent configurations)\footnote{These parameters 
are identical to the ones used in \cite{Fodor:2001au}}. From these two starting 
points with the use of the reweighting we predicted the plaquette 
($Pl$) and the $\bar{\psi}\psi$ expectation values and their uncertainties 
at $\beta=5.085$ and Im$(\mu)\neq 0$, that is at the target.
  In the following we refer to the technique 
starting from the phase transition point at Im$(\mu)=0$
as multi-parameter reweighting, 
while the technique starting from  $\beta=5.085$ (fixed at the target value) 
and Im$(\mu)= 0$ as Glasgow-type reweighting. We used the new Metropolis-type 
method defined above 
in both cases.  As a check for both starting points we also calculated the 
plaquette and the $\bar{\psi}\psi$ expectation values  directly by 
(\ref{multi-parameter2}) at the target points, and observed that the results 
are very similar to the results obtained by the 
Metropolis-type method.

The expectation values and their errors obtained by the new 
(Metropolis-type) method along with the direct simulation points are 
shown on the left panel of  Figure~\ref{fig:metro1}. 
When making the 
figure we had  to make use of an additional information, namely  the fact
  that the $\beta-$Im$(\mu)$
plane consists of three  sectors. If we want to make predictions 
about all of them based on the data at $\mu=0$ we can do it by explicitely 
using 
the Z(3) symmetry. That is we use the starting sample  in such 
a way that every plaquette value is included three times: 
once with the original weight and also with the weights of  shifted 
to $\mu \pm i\pi/N_t/3$ $\mu$ arguments.\footnote{Note that Z(3) symmetry 
is a specific feature of QCD at imaginary chemical potential, it has 
nothing to do with our new Metropolis-type method.}
Performing the reweighting like this 
we obtained Figure~\ref{fig:metro1}. 
In the left  panel  the 
second problem mentioned in the previous paragraph is seen. I.e. 
in case of Glasgow-type reweighting the small sample (1200 independent 
configurations) causes the problem. Namely, it does not contain any 
configuration belonging to the 
target phase when the target chemical potential is around 
Im($\mu)=\pi/12$. Then even our new method gives misleading results. 
Increasing the samples size, as soon as a single, dominant configuration 
turns up,  the new method supplies us with reliable error estimates.  
(See the right  panel of Figure~\ref{fig:metro1}, where we used 7000 
independent configurations in place of the 1200 configurations that 
lead to the wrong results on the left  panel.)
The right panel truly reflects
the correctness of the multi-parameter reweighting in contrast to
the single-parameter reweighting Glasgow- method.    
The infinitely large errors of the
Glasgow-method indicate that the whole
sample is thermalization, that is it does not provide information
about the expectation value in the corresponding imaginary chemical potential points. 

\begin{figure}[ht]
\begin{center}
\includegraphics*[width=7.0cm,bb=20 190 570 700]{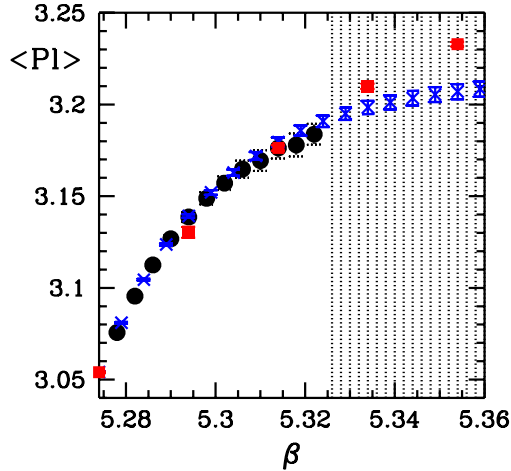}
\end{center}
\caption{\label{fig:pure_beta} {
Plaquette expectation values obtained by direct simulations and 
reweighting at various $\beta$-s. The squares and their errorbars 
denote the results of direct simulation. The initial sample used for 
reweighting consisted  of 33000 configurations. The 
parameters were: $\beta=5.274$, the critical $\beta$ in the $n_f=2+1$ 
flavour case when $m_{u,d}=0.096$, $m_s=2.08m_{u,d}$ are the quark 
masses (in lattice units) on a $4\cdot8^3$ size lattice. The circles 
and the errorbars belonging to them represent the newly introduced 
Metropolis-type method. Infinitely large errors mean that the whole, 
newly generated sample is thermalization so it does not contain 
information about the distribution at the $\beta$ parameter in question. 
The crosses  refer to the values given by the formula 
(5.2), the errors assigned to them are 
jackknife errors. It is clear  that the jackknife errors do not 
notice the limit of  validity of the reweighting procedure, but for 
small $|\Delta\beta|$-s they provide a good error estimate.    }}
\end{figure}   

An example of the  troubles  occuring when   using the jackknife error estimate 
will be shown 
in case of single-parameter reweighting in the
$\beta$ parameter. 
Thus one can use the new (expensive) error estimate to provide the 
limit of the applicability of the reweighting procedure after which the 
simpler jackknife method can be applied in the appropriate region. (This 
strategy was followed in the rest of the paper to calculate the EoS.) 
In our example we take   $4\cdot 8^3$ size 
lattice in case of $m_{u,d}=0.096$, $m_s=2.08m_{u,d}$ bare quark masses 
(in lattice units) at the critical point, i.e. at $\beta=5.274$. A sample 
simulated in this point consisting of 33000 configurations was reweighted 
by (\ref{multi-parameter2}). We evaluated the errors of the reweighted 
quantities by the Metropolis-type and by the jackknife methods. These are 
shown together with the reweighted plaquette expectation values in Figure 
\ref{fig:pure_beta}. It can be seen that the new method  gives 
information on the errors and also on the limit of the applicability of 
the reweighting procedure. On the other hand 
the jackknife error  can not be used outside a restricted 
 region and does not provide any clue on the region of applicability of the 
reweighting procedure.

\begin{figure}[ht]
\begin{center}
\includegraphics*[width=7.0cm,bb=20 190 570 700]{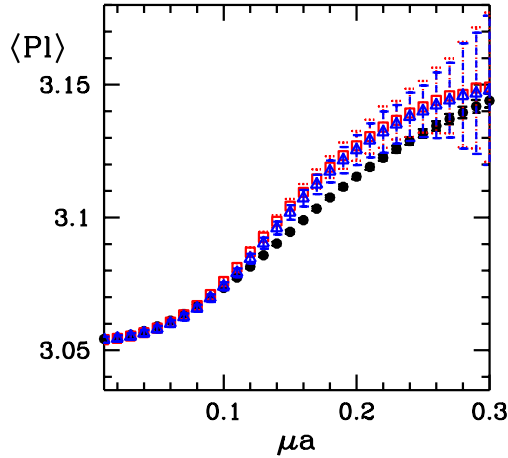}
\end{center}
\caption{\label{fig:metro} {
Average plaquette versus the real chemical potential. 
33000 configurations were simulated at the parameter set: $\beta=5.274$, 
$m_{u,d}=0.096$, $m_s=2.08m_{u,d}$ on a $4\cdot8^3$ size lattice. 
This  is at the critical $\beta$ in the $n_f=2+1$ flavour case. 
The circles  refer to the plaquette values 
given by  (5.2), their errorbars indicate jackknife errors. 
The squares belong to a Metropolis-type  method using the absolute values of 
the real part of the weights while in case of the triangles the absolute 
values of the weights were used. In both cases the errors 
are given by the square roots of the variances times the autocorrelation time. 
}}
\end{figure}

What can we say about the errors in the real $\mu$ case based on the new 
error estimation procedure? 
The Metropolis steps can only be taken using either  
the absolute values or the absolute values of the real parts of the weights. 
It is clear that both possibilities represent only approximations to the true reweighting given by eq.~(\ref{multi-parameter2}).
Whereas the expectation values provided by equation 
(\ref{multi-parameter2})  are exact in principle, we may  test  the above 
defined Metropolis-type procedures for the errors provided by them. 
We can claim based on Figure~\ref{fig:metro} that on a $4\cdot8^3$ size 
lattice in the $\mu \in [0,0.3]$ region and also  in all other examined 
cases the three methods do not provide significantly different expectation 
values, differences are tolerated in the errors. So we assign the errors 
of the Metropolis-type  method given through the use of the absolute values of 
the real parts of the weights to the expectation values obtained from  
equation~(\ref{multi-parameter2}). 

\section{Best reweighting lines and the LCP's}

\begin{figure}[ht]
\begin{center}
\includegraphics*[width=7.0cm,bb=20 170 570 700]{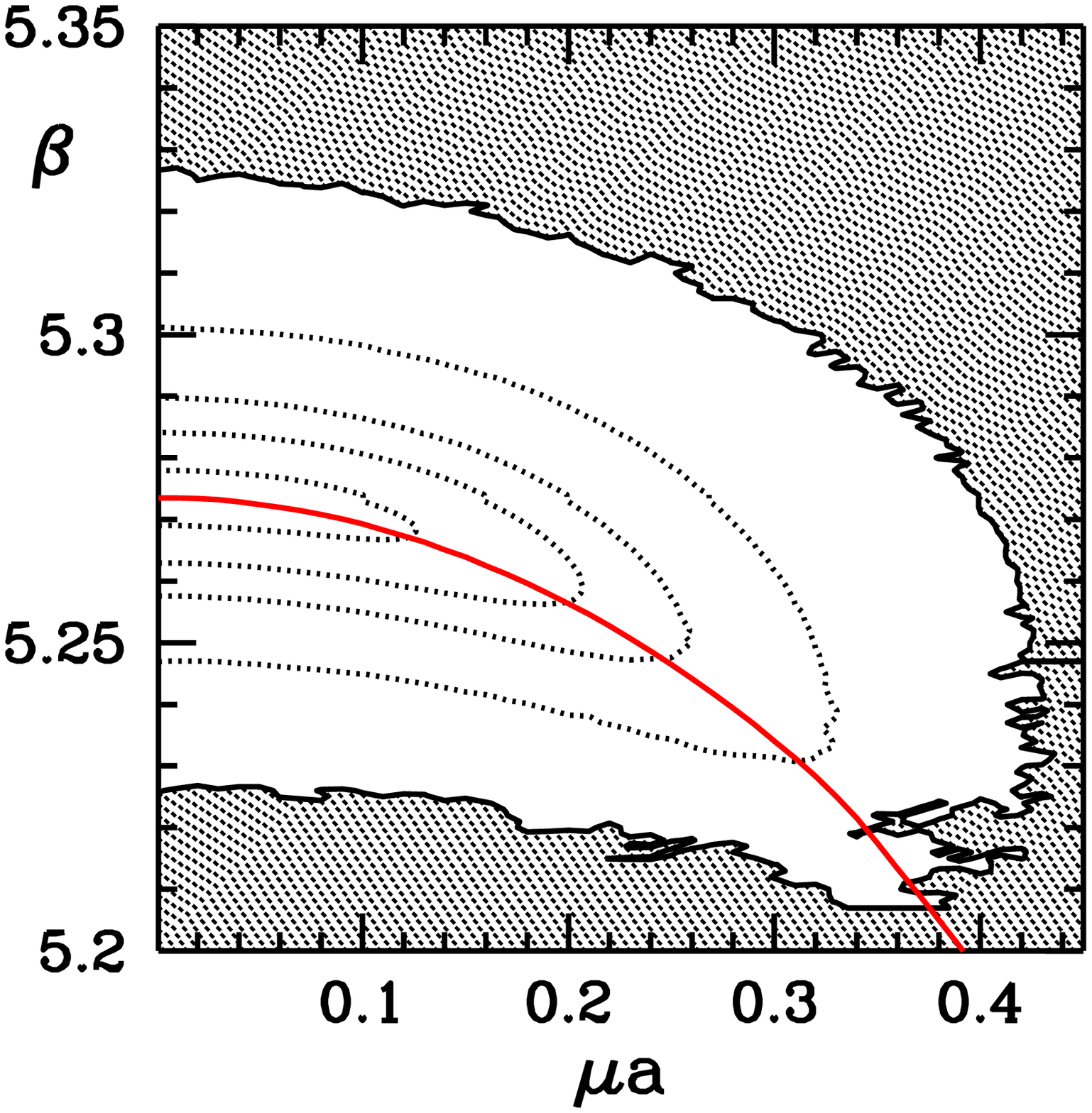}
\includegraphics*[width=7.0cm,bb=20 170 570 700]{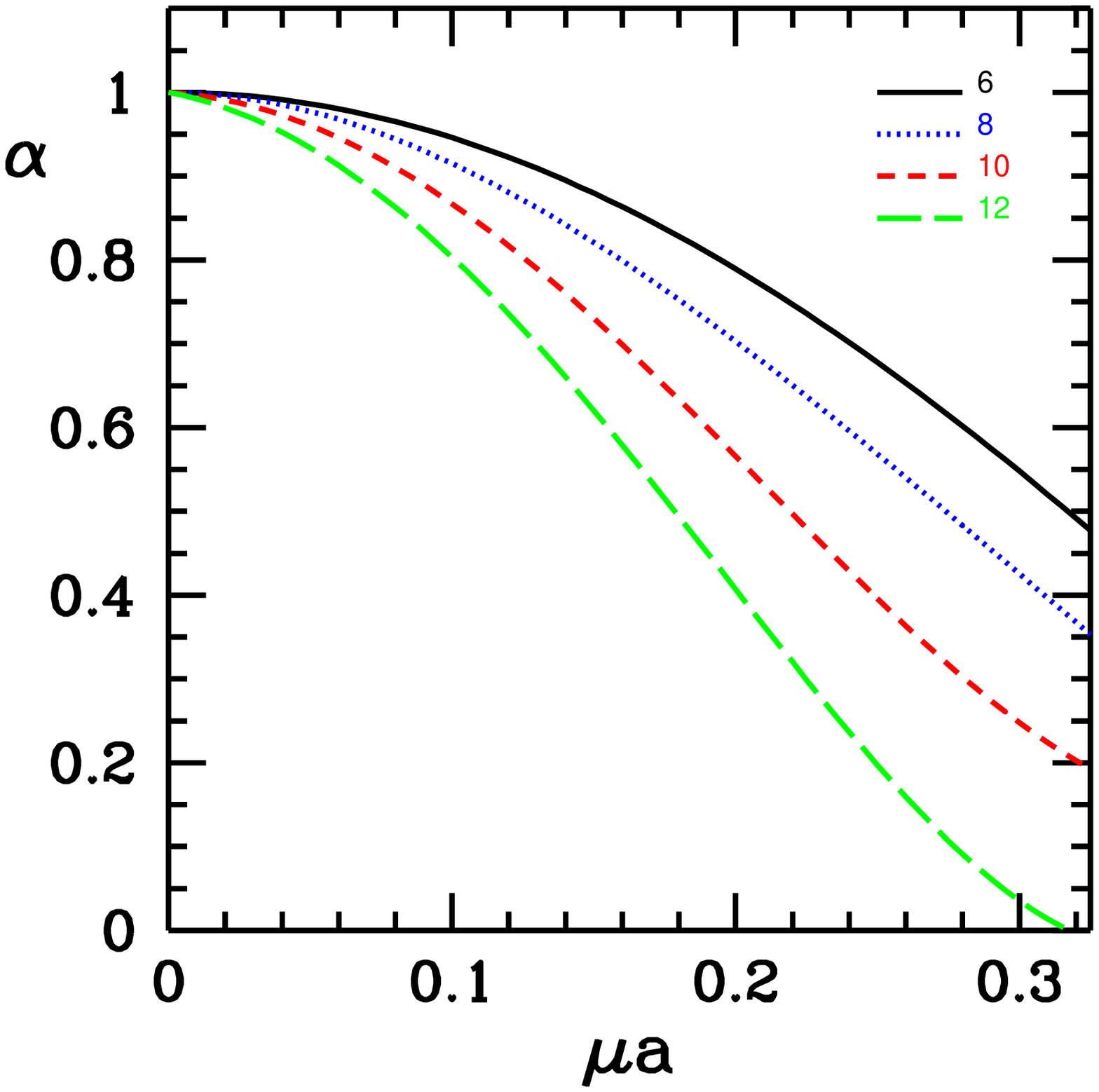}
\end{center}
\caption{\label{fig:contour} {
(a) The left panel shows the real chemical potential--$\beta$ plane.  
33000 configurations were simulated at the parameter set: $\beta=5.274$,
$m_{u,d}=0.096$, $m_s=2.08m_{u,d}$ on a $4\cdot8^3$ size lattice.
This  is at the critical $\beta$ in the $n_f=2+1$ flavour case. 
The dotted lines are the contours of the constant overlap. 
The dotted area is the unknown territory where the overlap vanishes, i.e. 
the sample regenerated from the initial one consists of nothing but 
thermalization. The roughness  of the border separating the unknown 
territory is due to the presence of random numbers in the procedure.
The solid line is the phase transition line determined by the peaks  
of susceptibility. (b) In the right panel the volume and the $\mu$ 
dependence of the overlap ($\alpha$) is shown. Upper curves correspond 
to smaller lattice sizes, $4\cdot 6^3$, $4\cdot 8^3$, $4\cdot 10^3$ and 
$4\cdot 12^3$ respectively. The half width ($\mu_{1/2}$; defined by 
  $\alpha\left.  \right|_{\mu_{1/2}}=1/2$) scales according to: 
$\mu_{1/2}\propto V^{-\gamma}$ with $\gamma\approx 1/3$. 
}}
\end{figure}

We can define reweighting lines on the $\beta$--$\mu$ 
plane so as to make the least possible mistake during the reweighting 
procedure. To do this we introduce the notion of overlap measure which 
we denote by $\alpha$. The overlap measure is the normalised number of 
different configurations in the sample created with the  
Metropolis-type 
reweighting after cutting off the thermalization. 
We plotted the contour lines of $\alpha$ in the left panel of 
Figure~\ref{fig:contour}. 
The dotted areas are 
unattainable, that means here the overlaps vanish, the errors are 
infinitely large.
The best reweighting lines can be defined for each simulation point. For a 
given value of $\mu$ we choose  $\beta$ corresponding to the maximal 
value of $\alpha$. The points of the best reweighting line are given by the 
rightmost points of the contours of constant overlap in  
Figure~\ref{fig:contour}~a. 

It is important to  examine the volume dependence of  reweighting by using 
the notion  of the overlap. The right panel of Figure~\ref{fig:contour} 
shows the $\mu$ dependence of the overlap at  fixed $\beta$ and  
quark mass parameters for different volumes ($V=4\cdot 6^3$, $4\cdot 
8^3$, $4\cdot 10^3$ and $4\cdot 12^3$). As expected, for fixed  $\mu$ 
larger volumes result in worse overlap. One can define the ``half-width'' 
($\mu_{1/2}$) of the $\mu$ dependence by the chemical potential value 
at which $\alpha=1/2$. One observes an approximate scaling behaviour 
for the half-width: $\mu_{1/2}\propto V^{-\gamma}$ with $\gamma \approx 1/3$, 
which is much better than a crude $1/\sqrt{V}$ estimate. 
Thus the volume dependence of $\mu- \beta$ reweighting is less restrictive 
than that of reweightings in other parameters. 
\begin{figure}[ht]
\begin{center}
\includegraphics*[width=13.5cm,bb=30 425 570 700]{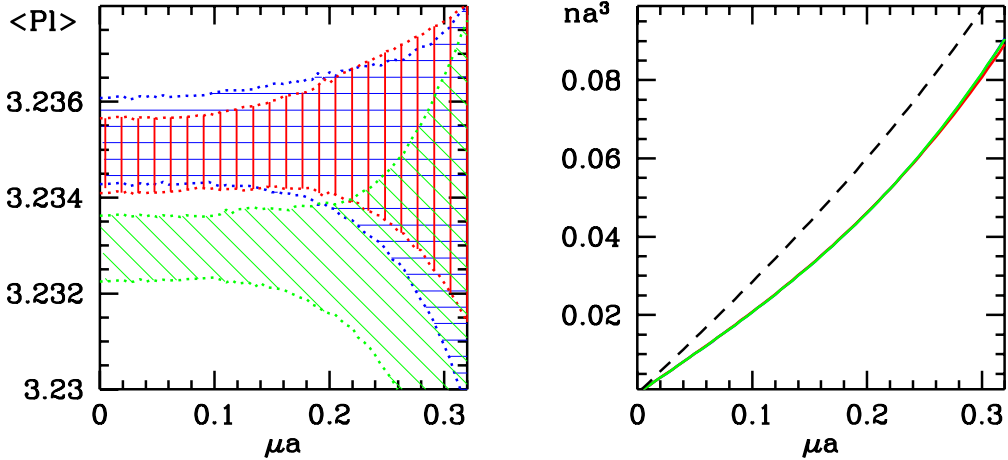}
\end{center}
\caption{\label{v-dependence} (a) The $\mu$ dependence of the
plaquette (left panel). Results obtained on $4\cdot 8^3$ lattices are
shown by horizontally, those on $4\cdot 10^3$ lattices are shown by vertically
and those on
$4\cdot 12^3$ are shown by diagonally lined bands. The results agree within
their statistical errors.  (b) The $\mu$ dependence of fermion number
density (right panel).  The difference of $4\cdot 8^3$, $4\cdot 10^3$
and $4\cdot 12^3$ results cannot be resolved (solid line). The dashed
line shows the Stefan-Boltzman limit. Simulations were done at
$\beta = 5.35$.}
\end{figure}

Expectation values of observables at finite chemical potential are
determined according to eq.~(\ref{multi-parameter2}).  
 There are quantities (e.g. plaquette,
Wilson-loop, Polyakov-line) for which the $\mu$ reweighting means only
a change in the weight of the configuration but not in the value of
the observable. For other quantites (e.g. chiral condensate, fermion
number) not only the weight but also the value is influenced by the
chemical potential since they are directly expressed by the fermion
matrix. It is instructive to study the volume dependence of both types of 
these quantities. The left panel of Figure \ref{v-dependence} shows the
average plaquette, the right panel the fermion number density, as
functions of the chemical potential. Both quantites are determined on
$4\cdot 8^3$, $4\cdot 10^3$ and $4\cdot 12^3$ lattices. The volume
dependence is practically negligible for the fermion number density,
whereas it is moderate for the average plaquette.

\begin{figure}[ht]
\begin{center}
\includegraphics*[width=6.5cm,bb=320 440 570 700]{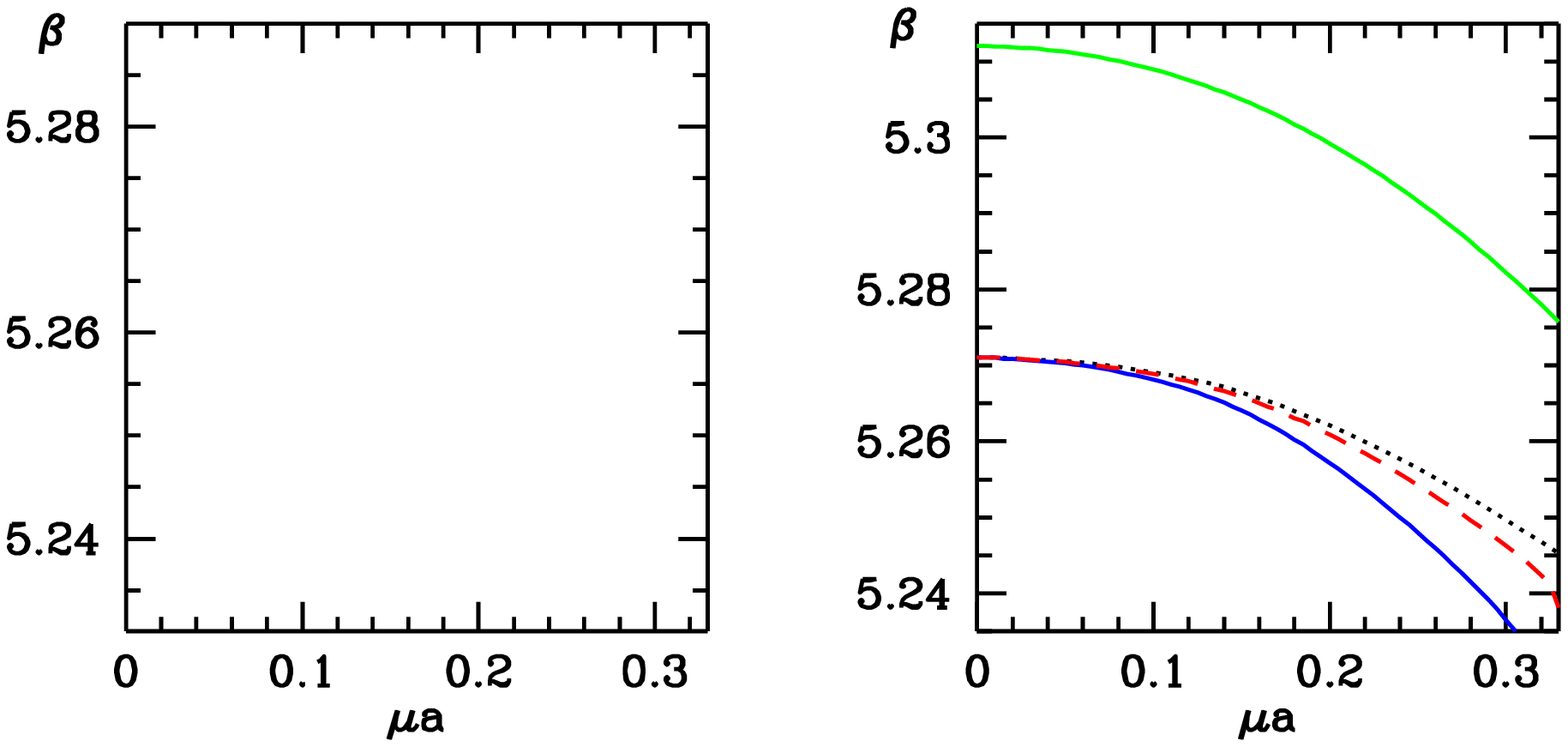}
\end{center}
\caption{ \label{reweight} Different types of best weight lines are shown 
on the $\mu$-$\beta$ plane. Three-parameter ($m$, $\beta$ and $\mu$) 
reweighting is used for the parameter set of LCP$_2(N_t=4)$ (dashed line).  
Along the line the system remains on the LCP, the small change in $\beta$ 
is compensated by the change of the mass parameter. Two parameter 
reweighting is used for the parameter set of LCP$_1(N_t=4)$ (upper solid line) 
and for that of LCP$_2(N_t=4)$ (lower solid line). The interpolating
technique is used to determine the line of constant physics on the
 $\mu$-$\beta$ plane (dotted line). Note that the results of the 
three-parameter reweighting technique and that of the interpolating 
technique agree within the statistical uncertainties.}
\end{figure}

It is obvious that the two-parameter reweighting used in the 
previous section does not follow  the LCP ($\beta$ gets smaller but 
the quark 
mass remains $m_0a$).  We show    in the following   that 
the best reweighting line along the LCP can be practically determined 
by two techniques: a.  three-parameter reweighting, b. 
 interpolating method.  

a. As it can be seen in the left panel of Figure \ref{fig:contour}, the
change in $\beta$ is not very large for the two-parameter reweighting.
Therefore, one can remain on the LCP by a simultaneous, small change
of the mass parameter of the lattice action. This results in a
three-parameter reweighting (reweighting in $ma$, $\beta$ and
$\mu a$). For some fixed $\mu_{target}$ two constraints are needed to 
determine $m_{target} a$ and $\beta_{target}$. One of them is the 
generalization of the new, Metropolis-type reweighting to some $m_{target} a\neq m_0 a$ 
and the other one is the LCP, which relates\footnote{In our specific case the 
$\beta_{LCP}(ma)$ relationship is given in Figure \ref{LCP}.} the mass 
parameter to the gauge coupling, i.e. one has to fulfill the following condition
\begin{equation}\label{unity3}
\beta_{target}=\beta_{LCP}(m_{target}a).
\end{equation}
Similarly to the two-parameter reweighting one can construct the 
best three-parameter reweighting line, which is shown in Figure~\ref{reweight}.

b. The other possibility to stay on the line of constant physics at finite
$\mu$ is the interpolating technique.  One uses the two-parameter
reweighting for two LCP's and interpolates between them. As we
discussed in Section 3 not only the mass parameter, but also $N_t$ (as
a continuous parameter) can be used to parametrize the LCP's. Thus the
data of Figure \ref{LCP} give $\beta_1(N_t,\mu=0)$ for LCP$_1$ and
$\beta_2(N_t,\mu=0)$ for LCP$_2$, respectively. The same $N_t$ values
for the two LCP's correspond to different mass parameters, which can
be written as $(am)_1=(am)_1(\beta_1,N_t,\mu=0)$ and
$(am)_2=(am)_2(\beta_2,N_t,\mu=0)$. Note that the two-parameter
reweighting to finite $\mu$ does not change the mass parameter.  Let
us take some fixed $N_t$ value and perform two-parameter reweightings
for both $\beta_1$ and $\beta_2$ parameter sets.  This results in
$\beta_1(\mu)$ and $\beta_2(\mu)$ best weight lines (these functions 
for $N_t$=4 are shown in Figure~\ref{reweight} by solid lines).  
Though these individual  curves leave the LCP-s,  
 there is a way to stay on e.g. LCP$_2$ by
interpolating between $\beta_1(\mu)$ and $\beta_2(\mu)$.  Since the
mass parameters of LCP$_1$ and LCP$_2$ are close to each other we use
a linear interpolation between them.  The obtained gauge coupling is
$\beta_i(\mu)$ and the mass parameter is $(am)_i(\mu)$.
\begin{eqnarray}\label{interpolate}
\beta_i(\mu)=\zeta \beta_1(\mu)+(1-\zeta)\beta_2(\mu),\\ \nonumber  
(am)_i(\mu)=\zeta (am)_1(\beta_1,N_t,\mu=0)+
(1-\zeta)(am)_2(\beta_2,N_t,\mu=0),\\ \nonumber 
\beta_i=\beta_2((am)_i).
\end{eqnarray}
The first two equations define the interpolation and the third
condition guarantees that we stay on LCP$_2$.  These three equations
determine the three unknown variables: $\beta_i$, $(am)_i$ and the
interpolating parameter $\zeta$.  The best weight line obtained by the
interpolation technique is also shown in Figure \ref{reweight}. 
As it can be seen the result of this method and the
predicition of the three-parameter reweighting agree quite well.  This
indicates that the requirement for the best overlap selects the same
weight lines even for rather different methods.

\section{Equation of state at non-vanishing chemical 
potential\label{sec_muneq0}}

In this section we study the EoS at finite chemical potential. Since
we are interested in the physics of finite baryon density we use
$\mu=\mu_u=\mu_d\neq 0$ for the two light quarks and $\mu_s=0$ for the
strange quark. The same technique and the same equations could be used
for the $\mu_s \neq 0$ case (one should simply add the strange quark
contribution to that of the light quarks).

The energy density and pressure can be derived in a similar way as in
the case of vanishing chemical potential discussed in Section \ref{sec_mu0}.
However, now the grand canonical potential ($\omega$) is used instead
of the free energy.  The definitions are:
\begin{eqnarray}
\epsilon (T,\mu)=\omega-T \frac{\partial{\omega}}{\partial {T}}
-\mu \frac{\partial{\omega}}{\partial{\mu}},\ \ \ \ \ \ \ \  &&
p(T,\mu) = -\omega. \ \ \ \ \ \ \ \ \
\end{eqnarray} 

Expressing the grand canonical potential in terms of the grand
canonical partition function ($\omega=-T/V\log Z_{gc}$) we get similar
results as in the $\mu=0$ case. In order to make this transparent we
explain the calculation in detail. As usual one has to separate the
space-like ($a_s$) and time-like ($a_t$) lattice parameters. This way
the derivatives with respect to $T$, $V$ and $\mu$ can be replaced by
derivatives with respect to $a_s$, $\xi$, and $\hat{\mu}$, where
$\xi=a_t/a_s$, $\hat \mu = \mu a_s$. It does not matter whether we write
$\mu a_t$ or $\mu a_s$, because we will set $\xi=1$ at the end of
calculation.  The new (lattice) variables can be expressed by the old
(thermodynamical) ones:
\begin{center}
\begin{tabular}{ccc}
$a_s=\frac{1}{N_s} V^{1/3},$\ \ \ \ \ \ \ \ \ \ &$ \xi=\frac{N_s}{N_t}
\frac{1}{T V^{1/3}}$, \ \ \ \ \ \ \ \ \ \  &$ \hat \mu = \frac{1}{N_s}
\mu V^{1/3}.$\ \ \ \ \ \ \ \ \ \ \\
\end{tabular}
\end{center}
And the derivatives:
\begin{eqnarray}
 \left. \frac{\partial}{\partial T}\right|_{V,\mu}&=&-\xi^2 N_t a_s  \left.\frac{\partial}{\partial
\xi}\right|_{a_s,\hat{\mu}},\nonumber \\
 \left. \frac{\partial}{\partial V}\right|_{T,\mu}&=& \frac{1}{3 V}\left( 
 a_s \left. \frac{\partial}{\partial
a_s}\right|_{\xi,\hat{\mu}} - \xi \left. \frac{\partial}{\partial \xi}
\right|_{a_s,\hat{\mu}} + \hat{\mu} \left. 
\frac{\partial}{\partial \hat{\mu}}\right|_{a_s,\xi} \right) \nonumber\\
\left. \frac{\partial}{\partial \mu}\right|_{T,V}&=& a_s \left. \frac{\partial}{\partial
\hat{\mu}}\right|_{a_s,\xi}. \nonumber
\end{eqnarray}
In particular, for  $\epsilon - 3p$ we have:
\begin{equation}
\epsilon-3p=\frac{T^2}{V} \left. \frac{\partial \log Z_{gc}}{\partial T}
\right|_{V,\mu}-3 T \left. \frac{\partial \log Z_{gc}}{\partial V}
\right|_{T,\mu}+\frac{\mu T}{V} \left. \frac{\partial \log
Z_{gc}}{\partial \mu} \right|_{V,T}. \nonumber
\end{equation}
Now we replace thermodynamical variables to lattice variables
\begin{eqnarray}
\epsilon-3p&=&-\frac{T}{V} \xi \left. \frac{\partial \log Z_{gc}}{\partial \xi}
\right|_{a_s,\hat{\mu}}+ \frac{\mu T}{V} a_s \left. \frac{\partial \log
Z_{gc}}{\partial \hat \mu} \right|_{a_s, \xi} \nonumber - \\ 
& &\frac{T}{V} \left(a_s \left. \frac{\partial
\log Z_{gc}}{\partial a_s} \right|_{\xi,\hat \mu}-\xi \left. \frac{\partial
\log Z_{gc}}{\partial \xi} \right|_{a_s, \hat \mu}+\hat {\mu}
\left. \frac{\partial \log Z_{gc}}{\partial \hat \mu} \right|_{a_s,
\xi} \right) \nonumber  \\
&=&-\frac{T}{V} a_s  \left. \frac{\partial\log Z_{gc}}{\partial a_s} \right|_{\xi,\hat \mu}.
\end{eqnarray}
 Setting $\xi=1$ we get:
\begin{equation}
\frac{\epsilon-3p}{T^4}=\left.-\frac{N_t^3}{N_s^3}a \frac{\partial (\log
Z_{gc})}{\partial a}\right|_{\mu a=const}.
\end{equation}

We can again write the derivative with respect to $a$ in terms of
derivatives with respect to the bare parameters. The additional bare
parameter $a \mu$ is kept constant, so it will not generate an extra
term:
\begin{equation}
\left.\frac{\partial}{\partial a}\right|_{\mu a=const}
=\left[\frac{\partial \beta}{\partial a} \frac{\partial}{\partial \beta}+
\sum_i \frac{\partial (a m_i)}{\partial a} \frac{\partial}{\partial (a m_i)}
\right]_{\mu a=const}.
\end{equation}
We have the same expression for $\epsilon-3p$ as in the $\mu=0$ case,
the only difference is that now all observables should be evaluated at
finite $\mu$:
\begin{equation}
\frac{\epsilon-3p}{T^4}=-N_t^4 a 
\left(\overline{{\rm Pl}}^{(\mu)}\left. 
\frac{\partial \beta}{\partial a}\right|_{\rm LCP}+
\overline{\bar{\Psi}\Psi}_{ud}^{(\mu)} m_{ud} + \overline{\bar{\Psi}\Psi}_{s}^{(\mu)} m_s \right).
\end{equation}

\begin{figure}[ht]
\begin{center}
\epsfig{file=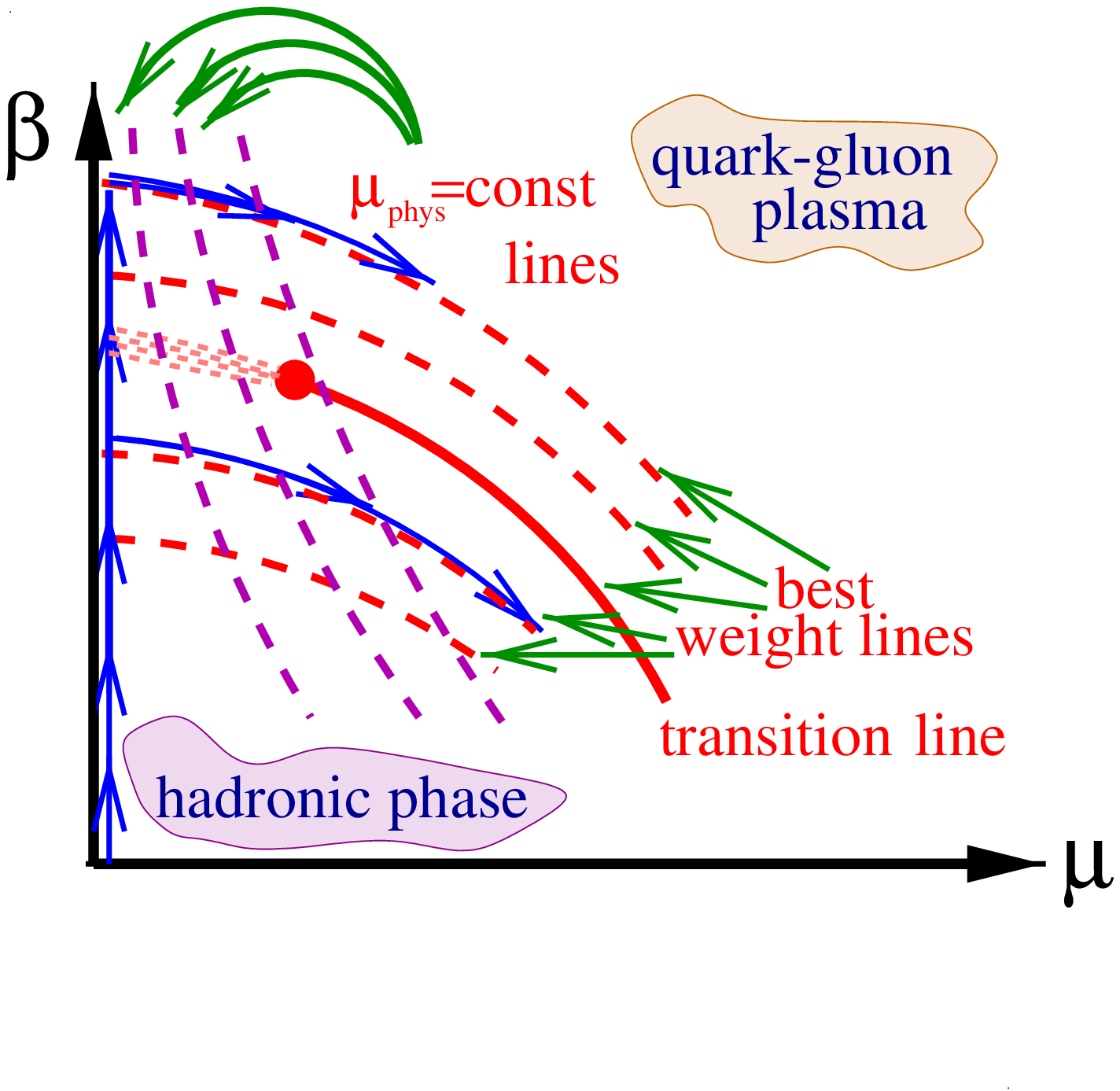, height=7.5cm, width=7.5cm}
\end{center}
\caption{ \label{tmu}  Illustration of the integral method at finite chemical
potential (left panel). The solid lines are $\mu={\rm const}$ lines on
the $\mu a - \beta$ plane. Dashed lines are the best reweighting lines
starting from different simulation points. Arrows show the path of
integration we used when evaluating eq.~(7.7). }
\end{figure}

For the pressure we have an additional variable in the integral, the
gradient of $\log Z_{gc}$ has now an extra component in the $a \mu$
direction:
\begin{equation}
\frac{p}{T^4}=-\frac{N_t^3}{N_s^3}
\int^{(\beta, a m_i,a \mu)}_{(\beta_0,a m_{i0}, a\mu=0)}
d (\beta,a m_i,a \mu) 
\left(\begin{array}{c}
{\partial \log Z_{gc}}/{\partial \beta} \\
{\partial \log Z_{gc}}/{\partial (a m_i)} \\
{\partial \log Z_{gc}}/{\partial (a \mu)} 
\end{array} \right )- \frac{p_0}{T^4}.
\end{equation}
The subtracted $p_0$ term is now the pressure at $T=0$ and $\mu=0$,
i.e. the same as before. We can rewrite the above equation in terms of
observables:
\begin{eqnarray}
\frac{p}{T^4}&=&
-N_t^4\int^{(\beta,a m_i, a\mu)}_{(\beta_0,a m_{i0},a\mu=0)}
d (\beta,a m_{ud},a m_s,a\mu) 
\left(\begin{array}{ll}
&\langle{\rm Pl}\rangle^{(\mu)} \\
&\langle\bar{\Psi}\Psi_{ud}\rangle^{(\mu)} \\
&\langle\bar{\Psi}\Psi_{s}\rangle^{(\mu)} \\
{N_s^3N_t}^{-1}&\langle{\frac{\partial \log \det M}{\partial (a\mu)}\rangle^{(\mu)}}
\end{array} \right) \nonumber \\
\label{eq:pt4}
\end{eqnarray}
where the superscript $\mu$ denotes the expectation values of the
observables at reweighted $\mu$ values:
\begin{equation}
\langle  {\cal O}(\beta,\mu,m) \rangle^{(\mu)}= 
{\overline {{\cal {O}}}(\beta,\mu,m)}_{T\neq 0}-
{\overline {{\cal O}}(\beta,\mu=0,m)}_{T=0}.
\end{equation}

The result should again be independent of the integration path. We
chose the following paths (cf. Fig.  \ref{tmu}). First we integrated from ($\beta_0$, $a
m_{i0}$) to some ($\beta_1$, $a m_{i1}$) along the LCP with $\mu=0$. Then
we followed the line of best reweighting to reach the required
($\beta$, $am_i$, $a\mu$) point.  We  checked that the result remains
unchanged if we do it in the opposite way, i.e. we go first along the
best reweighting line and then along the constant $a \mu$ line. The
results along the two integration lines were equal within statistical
errors.

\begin{figure}[ht]
\begin{center}
\epsfig{file=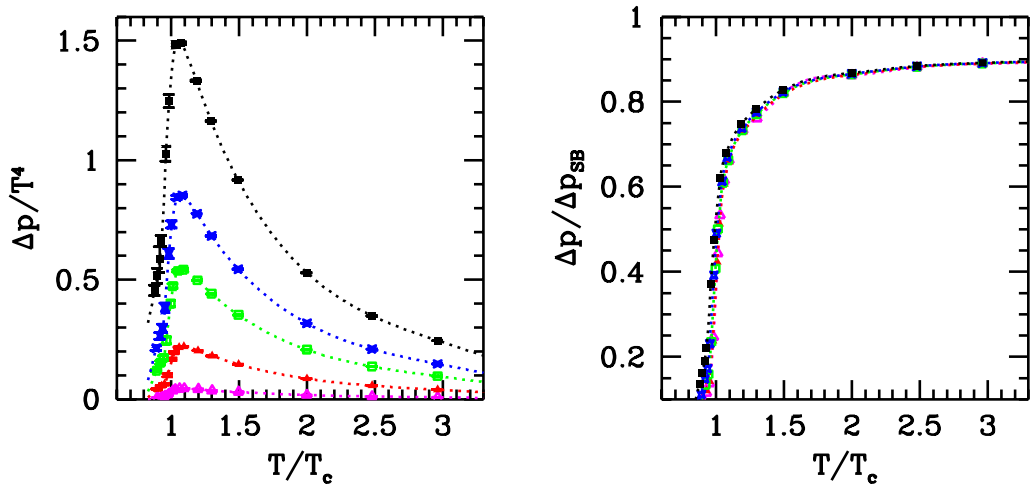,height=7.5cm, width=15.0cm}
\end{center}
\caption{\label{eosmu} The equation of state at,
$\mu_B =100$, 210, 330, 410 and 530~MeV. The left panel shows
the pressure difference between the $\mu = 0$ and $\mu\neq 0$ cases
normalized by $T^4$, whereas on the right panel the normalization is
done by the $N_t=4$ lattice Stefan-Boltzman limit (in continuum
this limit is
$n_f(\mu/T)^2$). Note that $\Delta p / \Delta p^{SB}$ seems to show
some scaling behaviour (it depends mostly on the temperature; whereas
its dependence on $\mu$ is much weaker). Thus, the $\mu$ dependence of
$\Delta p$ is almost completely given by the $\mu$ dependence of the
free gas.  }
\end{figure}

\begin{figure}[ht]
\begin{center}
\epsfig{file=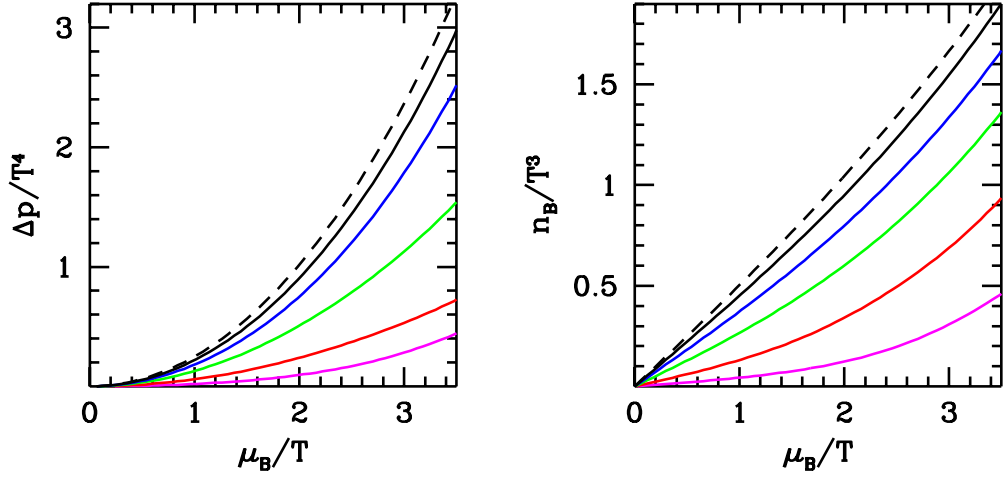,height=7.5cm, width=15.0cm}
\end{center}
\caption{(a) The difference of the pressure at finite $\mu_B$ and
$\mu_B=0$ (left panel) as a function of $\mu_B/T$ at $T/T_c=0.9$, 0.98,
 1.03, 1.2, 2.28.  (b) Dimensionless baryon number density as a
function of $\mu_B$ at $T/T_c= 0.9$, 0.98, 1.03, 1.2, 2.48. Dashed
line shows the SB limit in both cases. The higher the temperature is,
the closer the lines run to the SB limit.\label{eosmu2} }
\end{figure}

\begin{figure}[ht]
\begin{center}
\epsfig{file=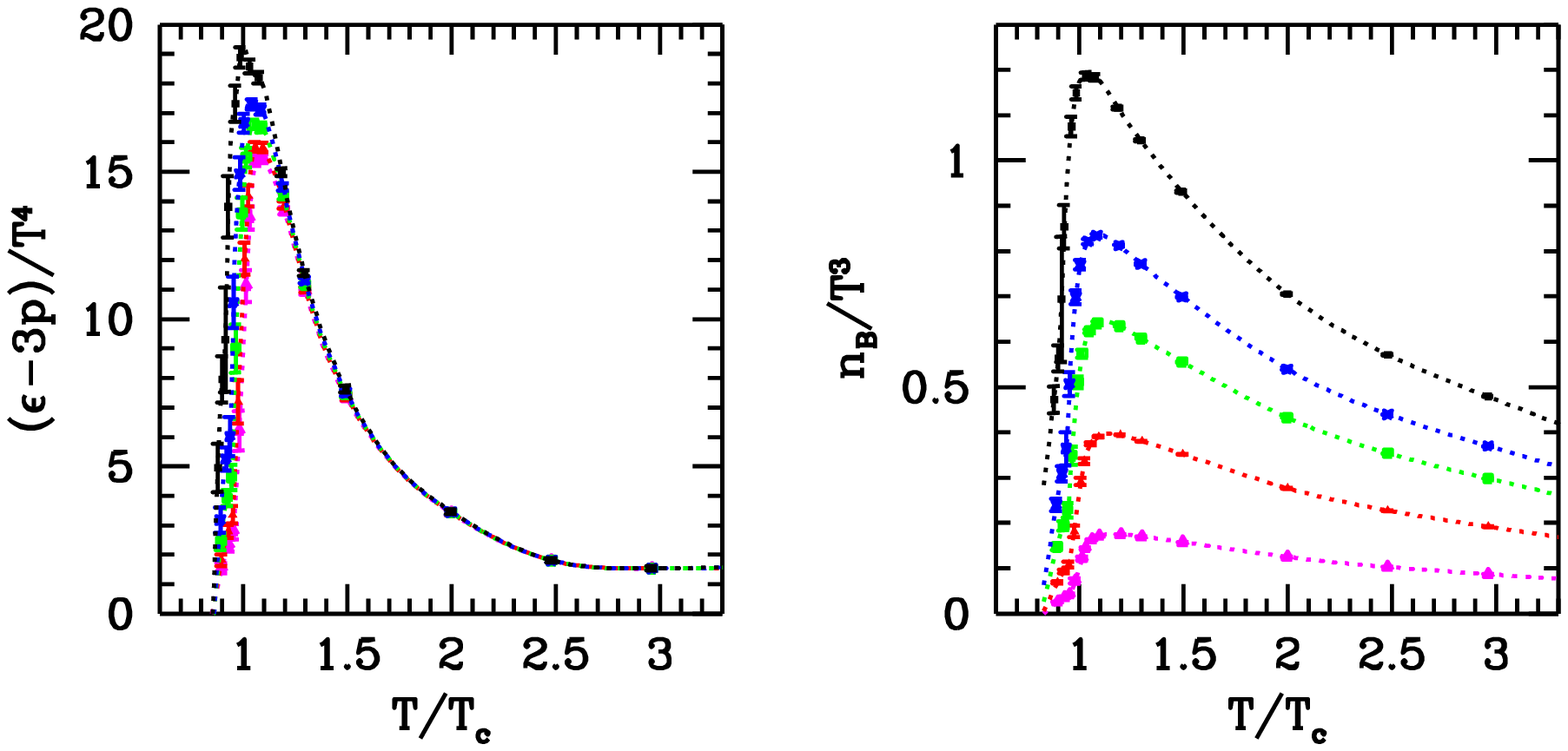,height=7.5cm, width=15.0cm}
\end{center}
\caption{\label{eosmu3}
(a) $(\epsilon-3p)/T^4$ (b) Dimensionless baryon
number density as a function of $T/T_c$ at $\mu_B$=100,210,330,410~MeV and
$\mu_B$=530~MeV.
}
\end{figure}

We present lattice results on $\Delta p(\mu,T)=p(\mu\neq
0,T)-p(\mu=0,T)$, $\epsilon(\mu,T)$-3$p(\mu,T)$ and $n_B(\mu,T)$. Our
statistical errorbars are also shown. They are rather small, in many
cases they are even smaller than the thickness of the lines.

On the left panel of Fig. \ref{eosmu} we present $\Delta p/T^4$ for
five different $\mu$ values. On the right panel normalisation is done
by $\Delta p^{SB}$, which is $\Delta p(\mu,T\rightarrow\infty)$.
Notice the interesting scaling behaviour.  $\Delta p / \Delta p^{SB}$
depends only on T and it is practically independent from $\mu$ in the
analysed region.  The left panel of Fig. \ref{eosmu2} shows the same
pressure difference as a function of $\mu _B /T$ for five different
temperatures. The right panel shows dimensionless baryon number density
($n_B/T^3$) as a function of $\mu _B /T$ at the same temperatures.
The left panel of Fig. \ref{eosmu3} shows $\epsilon$-3$p$ normalised by
$T^4$, which tends to zero for large $T$.  The right panel of
Fig. \ref{eosmu3} gives the dimensionless baryonic density as a
function of $T/T_c$ for different $\mu$-s.

Here we summarise what kind of errors occur during our lattice
calculations. The  error coming from reweighting has been amply 
discussed in Section 5.
Another source of error is the finiteness of the physical
volume. The volume dependence of physical observables
 is smaller than the statistical errors for the
plaquette average or  quark number density. Systematic errors
coming from non-zero microcanonical step-size cause a 0.2\% and 0.1\%
error in the value of physical observables for zero temperature and
finite temperature calculations, respectively.

\section{Conclusions, outlook}

In this paper we studied the thermodynamical properties of QCD at
finite chemical potential $\mu$. We used the overlap enhancing
multi-parameter reweighting method proposed by two of us
\cite{Fodor:2001au} and its generalization to $2+1$ flavour staggered
QCD \cite{Fodor:2002pe}. Our primary goal was to determine the
equation of state (EoS) on the line of constant physics (LCP) at
finite temperature and chemical potential.

We have pointed out that even at $\mu$=0 the EoS depends on the fact
whether we are on an LCP or not. Note that previous results in the
staggered formalism usually used the ``non-LCP approach''.   According to our
findings pressure and $\epsilon$ - 3p (interacion measure) 
on the LCP has different high
temperature behaviour than in the ``non-LCP approach''. 

We discussed the reliability of the reweighting technique. We introduced 
an error estimate, which successfully shows the limits of the method 
yielding infinite errors in the parameter regions, where reweighting 
gives wrong results. 
We showed how to define and determine the best
weight lines on the $\mu$--$\beta$ plane.  

We discussed the two-parameter and three-parameter reweighting
techniques. Two techniques were
presented (three-parameter reweighting and the interpolating method)
to stay on the LCP even when reweighting to non-vanishing chemical
potentials.

We calculated the thermodynamic equations for $\mu \neq 0$ and
determined the EoS along an LCP. We presented lattice data on the 
pressure, the interaction-measure and the baryon number density as a function
of temperature and chemical potential. The physical range of our
analysis extended upto $500-600$~MeV in temperature and baryon chemical
potential as well.

Clearly much more work is needed to get the final form of
non-perturbative EoS of QCD. One has to extrapolate to zero step-size
in the R-algorithm. Extrapolation to the thermodynamic and continuum
limits is a very CPU demanding task in the $\mu \neq 0$
case.  Physical
$m_\pi/m_\rho$ ratio should be reached by decreasing the light quark
mass. Finally, renormalised LCP's (LCP$^*$ in our notation) should be
used when evaluating thermodynamic quantites.

\section{Acknowledgements}

This work was partially supported by Hungarian Scientific
grants, OTKA-T37615/\-T34980/\-T29803/\-M37071/\-OMFB1548/\-OMMU-708. 
For the simulations a modified version of the MILC
public code was used (see http://physics.indiana.edu/\~{ }sg/milc.html). 
The simulations were carried out on the 
E\"otv\"os Univ., Inst. Theor. Phys. 163 node parallel PC cluster.

\end{document}